\documentclass[times,tighten,twocolumn,twocolappendix,resetfootnote]{aastex7}

\usepackage{graphicx}
\usepackage{txfonts}
\usepackage{lipsum}
\usepackage{longtable}  
\usepackage{float}
\usepackage{xcolor}

\newcommand{\xmm}{{\em XMM-Newton}}

\newcommand{\swift}{{\em Swift}}

\newcommand{\nustar}{{\em NuSTAR}}

\newcommand{\nicer}{{\em NICER}}
\newcommand{\ixpe}{{\em IXPE}}

\newcommand{\be}{\begin{equation}}
\newcommand{\en}{\end{equation}}
\def\ltsima{$\; \buildrel < \over \sim \;$}
\def\lsim{\lower.5ex\hbox{\ltsima}}
\def\gtsima{$\; \buildrel $\geq$ \over \sim \;$}
\def\gsim{\lower.5ex\hbox{\gtsima}}

\def\arc{\mbox{$^{\prime\prime}$}}

\def\deg{\mbox{$^{\circ}$}}

\def\lum {\mbox{erg~s$^{-1}$}}

\shorttitle{Multiband polarimetry of PSR J1023+0038}
\shortauthors{Baglio, Coti Zelati et al.}

\received{\today}
\revised{\today}
\accepted{\today}

\begin{document}

\title{Polarized multiwavelength emission from pulsar wind -- accretion disk interaction in a transitional millisecond pulsar}

\correspondingauthor{M.~C.~Baglio, F.~Coti~Zelati}

\author[orcid=0000-0003-1285-4057]{Maria~Cristina~Baglio}
\affiliation{INAF--Osservatorio Astronomico di Brera, Via Bianchi 46, I-23807 Merate (LC), Italy}
\email[show]{maria.baglio@inaf.it}
\thanks{The first two authors equally contributed to this work.}

\author[orcid=0000-0001-7611-1581]{Francesco~Coti~Zelati}
\affiliation{Institute of Space Sciences (ICE, CSIC), Campus UAB, Carrer de Can Magrans s/n, E-08193 Barcelona, Spain}
\affiliation{Institut d'Estudis Espacials de Catalunya (IEEC), 08860 Castelldefels (Barcelona), Spain}
\affiliation{INAF--Osservatorio Astronomico di Brera, Via Bianchi 46, I-23807 Merate (LC), Italy}
\email[show]{cotizelati@ice.csic.es}
\thanks{The first two authors equally contributed to this work.}

\author[orcid=0000-0003-0331-3259]{Alessandro~Di~Marco}
\affiliation{INAF Istituto di Astrofisica e Planetologia Spaziali, Via del Fosso del Cavaliere 100, I-00133 Rome, Italy}
\email{alessandro.dimarco@inaf.it}

\author[orcid=0000-0001-8916-4156]{Fabio~La~Monaca}
\affiliation{INAF--Istituto di Astrofisica e Planetologia Spaziali, Via del Fosso del Cavaliere 100, I-00133 Rome, Italy}
\affiliation{Dipartimento di Fisica, Universit\`{a} degli Studi di Roma ``Tor Vergata'', Via della Ricerca Scientifica 1, I-00133 Roma, Italy}
\email{fabio.lamonaca@inaf.it}

\author[orcid=0000-0001-6289-7413]{Alessandro~Papitto}
\affiliation{INAF--Osservatorio Astronomico di Roma, Via Frascati 33, I-00078 Monteporzio Catone (RM), Italy}
\email{alessandro.papitto@inaf.it}

\author[orcid=0000-0003-0764-0687]{Andrew~K.~Hughes}
\affiliation{Department of Physics, University of Alberta, Edmonton, T6G 2E1, Canada}
\affiliation{Astrophysics, Department of Physics, University of Oxford, Keble Road, Oxford, OX1 3RH, UK}
\email{hughes1@ualberta.ca}

\author[orcid=0000-0001-6278-1576]{Sergio~Campana}
\affiliation{INAF--Osservatorio Astronomico di Brera, Via Bianchi 46, I-23807 Merate (LC), Italy}
\email{sergio.campana@inaf.it}

\author[orcid=0000-0002-3500-631X]{David~M.~Russell}
\affiliation{Center for Astrophysics and Space Science, New York University Abu Dhabi, PO Box 129188, Abu Dhabi, UAE}
\email{dave.russell@nyu.edu}

\author[orcid=0000-0002-1522-9065]{Diego~F.~Torres}
\affiliation{Institute of Space Sciences (ICE, CSIC), Campus UAB, Carrer de Can Magrans s/n, E-08193 Barcelona, Spain}
\affiliation{Institut d'Estudis Espacials de Catalunya (IEEC), 08860 Castelldefels (Barcelona), Spain}
\affiliation{Institució Catalana de Recerca i Estudis Avançats (ICREA), Passeig Lluís Companys 23, E-08010 Barcelona, Spain}
\email{dtorres@ice.csic.es}

\author[orcid=0000-0002-0426-3276]{Francesco~Carotenuto}
\affiliation{INAF--Osservatorio Astronomico di Roma, Via Frascati 33, Monteporzio Catone (RM), I-00078, Italy}
\affiliation{Astrophysics, Department of Physics, University of Oxford, Keble Road, Oxford, OX1 3RH, UK}
\email{francesco.carotenuto@inaf.it}

\author[orcid=0000-0001-9078-5507]{Stefano~Covino}
\affiliation{INAF--Osservatorio Astronomico di Brera, Via Bianchi 46, I-23807 Merate (LC), Italy}
\affiliation{Como Lake centre for AstroPhysics (CLAP), DiSAT, Universit\`{a} dell’Insubria, via Valleggio 11, 22100 Como, Italy}
\email{stefano.covino@inaf.it}

\author[orcid=0000-0002-5069-4202]{Domitilla~de~Martino}
\affiliation{INAF--Osservatorio Astronomico di Capodimonte, Salita Moiariello 16, I-80131 Naples, Italy}
\email{domitilla.demartino@inaf.it}

\author[orcid=0000-0002-2815-7291]{Stefano~Giarratana}
\affiliation{INAF--Osservatorio Astronomico di Brera, Via Bianchi 46, I-23807 Merate (LC), Italy}
\email{stefano.giarratana@inaf.it}

\author[orcid=0000-0002-6154-5843]{Sara~E.~Motta}
\affiliation{INAF--Osservatorio Astronomico di Brera, Via Bianchi 46, I-23807 Merate (LC), Italy}
\email{sara.motta@inaf.it}

\author[orcid=0000-0003-0168-9906]{Kevin~Alabarta}
\affiliation{Center for Astrophysics and Space Science, New York University Abu Dhabi, PO Box 129188, Abu Dhabi, UAE}
\email{kalabarta@nyu.edu}

\author[orcid=0000-0001-7164-1508]{Paolo~D'Avanzo}
\affiliation{INAF--Osservatorio Astronomico di Brera, Via Bianchi 46, I-23807 Merate (LC), Italy}
\email{paolo.davanzo@inaf.it}

\author[orcid=0000-0003-4795-7072]{Giulia~Illiano}
\affiliation{INAF--Osservatorio Astronomico di Roma, Via Frascati 33, I-00078 Monteporzio Catone (RM), Italy}
\email{giulia.illiano@inaf.it}

\author{Marco M.~Messa}
\affiliation{Dipartimento di Fisica, Università degli Studi di Milano, Via Celoria 16, I-20133, Milan, Italy}
\affiliation{INAF--Osservatorio Astronomico di Brera, Via Bianchi 46, I-23807 Merate (LC), Italy}
\email{marco.messa@inaf.it}

\author[orcid=0000-0002-0943-4484]{Arianna~Miraval~Zanon}
\affiliation{ASI - Agenzia Spaziale Italiana, Via del Politecnico snc, I-00133 Rome, Italy}
\email{arianna.miraval@inaf.it}

\author[orcid=0000-0003-2177-6388]{Nanda~Rea}
\affiliation{Institute of Space Sciences (ICE, CSIC), Campus UAB, Carrer de Can Magrans s/n, E-08193 Barcelona, Spain}
\affiliation{Institut d'Estudis Espacials de Catalunya (IEEC), 08860 Castelldefels (Barcelona), Spain}
\email{rea@ice.csic.es}


\begin{abstract}
Transitional millisecond pulsars (tMSPs) bridge the evolutionary gap between accreting neutron stars in low-mass X-ray binaries and millisecond radio pulsars. These systems exhibit a unique subluminous X-ray state characterized by the presence of an accretion disk and rapid switches between high and low X-ray emission modes. The high mode features coherent millisecond pulsations spanning from the X-ray to the optical band. We present multiwavelength polarimetric observations of the tMSP PSR\,J1023$+$0038 aimed at conclusively identifying the physical mechanism powering its emission in the subluminous X-ray state. During the high mode, we report a probable detection of polarized emission in the 2--6\,keV energy range, with a polarization degree of $(12 \pm 3)\%$ and a polarization angle of $-2^\circ \pm 9^\circ$ measured counterclockwise from the North celestial pole towards the East (99.7\% c.l.; uncertainties are quoted at 1$\sigma$). At optical wavelengths, we find a polarization degree of $(1.41 \pm 0.04)\%$ and a polarization angle aligned with that in the X-rays, suggesting a common physical mechanism operating across these bands. Remarkably, the polarized flux spectrum matches the pulsed emission spectrum from optical to X-rays. The polarization properties differ markedly from those observed in other accreting neutron stars and isolated rotation-powered pulsars and are also inconsistent with an origin in a compact jet. Our results provide direct evidence that the polarized and pulsed emissions both originate from synchrotron radiation at the boundary region formed where the pulsar wind interacts with the inner regions of the accretion disk.
\end{abstract}



\section{Introduction}
\label{sec:intro}
The evolution of accreting neutron stars (NSs) in low-mass X-ray binaries (LMXBs) as progenitors of millisecond radio pulsars (MSPs) involves distinct phases \citep{Alpar1982}, with transitional MSPs (tMSPs) representing the most intriguing stage of this process due to their unique properties \citep{campana18b, papitto22}. PSR\,J1023$+$0038 (J1023) is the prototypical tMSP \citep{Archibald2009}, offering an opportunity to study these enigmatic objects in detail. Detected in 2007 as a radio MSP with a spin period of 1.69\,ms, J1023 orbits a 0.24 $M_{\odot}$ star every 4.75 hours \citep{Archibald2009}. 

In 2013, J1023 experienced dramatic changes in its emission properties, ceasing its radio pulsations and exhibiting increased emission across the electromagnetic spectrum \citep{Stappers2014,Patruno2014}.
J1023 has since remained in an active state, characterized by subluminous X-ray emission with an average luminosity of $\simeq$$7 \times 10^{33}$\,\lum\ in the 0.3--80\,keV energy range \citep{cotizelati18a} -- over four orders of magnitude below the Eddington luminosity for a NS. This emission alternates between high-intensity (70--80\% of the time) and low-intensity (20--30\%) modes, with occasional brief flares reaching $\approx$10$^{35}$\,\lum\ \citep{Papitto13,Linares2014ApJ,Archibald15,Bogdanov15}. Mode switches occur within 10--30\,s, with low modes lasting from tens of seconds to minutes. During high modes, which typically last from minutes to several tens of minutes, J1023 exhibits pulsed emission at the pulsar spin period at X-ray, ultraviolet (UV), and optical wavelengths, pointing to processes linked to the NS rotation \citep{Archibald15, papitto19, miravalzanon22, illiano23}. 

This pulsed emission disappears during the low mode. The UV, optical, and near-infrared emissions mainly originate from an accretion disk surrounding the NS and the irradiated companion star, leading to flickering and flaring \citep{kennedy18, papitto18, hakala18, shahbaz18}. Bright radio--mm continuum emission is observed, with increased radio emission during the low mode and brief mm-band flares during switches from high to low modes \citep{deller15, bogdanov18, Baglio2023}.

The behavior of J1023 raises fundamental questions about the mechanisms powering its emission in the active state. J1023 releases rotational kinetic energy at a rate of $\simeq4.4 \times 10^{34}$\,\lum\ \citep{archibald13}, about six times its average X-ray luminosity. This has sparked debates on whether the emission of tMSPs like J1023 in this active state is powered by pulsar rotation, mass accretion, or a combination of both \citep{bogdanov18,papitto19,Veledina19,linares22}. Multiband polarization measurements provide a diagnostic tool for investigating these processes and solving this enigma.

This Letter presents the results from the first multiwavelength polarimetric campaign of J1023 and is structured as follows. Section\,\ref{sec:obs} describes the observations and data processing techniques employed across the X-ray, optical, and radio bands using the \emph{Imaging X-ray Polarimetry Explorer} (\ixpe), the \emph{Neutron Star Interior Composition Explorer Mission} (\nicer), the \emph{Neil Gehrels Swift Observatory} (\swift), the \emph{Very Large Telescope} (VLT), and the \emph{Karl G. Jansky Very Large Array} (VLA). Section\,\ref{sec:data_analysis} details the data analysis and results, encompassing the selection of emission modes and timing analysis of the \ixpe\ data, and the polarimetric analysis across different wavelengths. We also present the spectral energy distributions (SEDs) derived from our observations. Section\,\ref{sec:discussion} discusses the implications of our findings on the emission mechanisms powering J1023 during its active state, evaluating scenarios such as accretion-powered processes, rotation-powered mechanisms analogous to isolated pulsars, and interactions between the pulsar wind and the accretion flow. Section\,\ref{sec:conclusions} summarizes how our multiwavelength polarimetric observations provide critical insights into the enigmatic behavior of tMSPs like J1023.

\section{The multi-wavelength dataset}
\label{sec:obs} 
We acquired multiwavelength data of J1023 using space-borne and ground-based telescopes (see Table\,\ref{tab:log}). This section describes the observations and the procedures used for data processing and analysis. All timestamps are referenced to the Solar System barycenter using the JPL DE440 ephemeris \citep{Park2021} and the position: RA = 10$^\mathrm{h}$23$^\mathrm{m}$47$^\mathrm{s}$69, Dec = +00$^{\circ}$38$^{\prime}$40.$^{\prime\prime}$8 (J2000.0; \citealt{Deller2012}).

\begin{table*}
\footnotesize
\caption{Journal of the observations of J1023.}
\label{tab:log}
\centering
\begin{tabular}{lccccc}
\hline\hline
Telescope/Instr.& Obs./Prg. Id    & Setup         & Start -- End time	            & Exposure & Band \\
			&                 &               & YYYY Mmm DD hh:mm:ss (UTC)	    & (ks)   &    \\
\hline
\ixpe/DUs             & 03005599         &                  & 2024 May 29 12:30:51  -- 2024 Jun 14 14:57:45  & 675 & 2--8\,keV    \\
\nicer/XTI        & 7034060107       &                  & 2024 Jun 01 02:42:20 -- 2024 Jun 01 21:36:39  & 4.3  & 0.5--10\,keV    \\
\swift/XRT/UVOT  & 00033012239     & PC / Event mode  & 2024 Jun 04 22:14:59 -- 2024 Jun 04 23:59:54  & 1.5  & 0.3--10\,keV / $UVM2$    \\
\swift/XRT/UVOT  & 00033012240     & PC / Event mode  & 2024 Jun 05 01:23:09 -- 2024 Jun 05 01:48:53  & 1.5  & 0.3--10\,keV / $UVM2$    \\
VLA             & 24A-476         & B configuration   & 2024 Jun 04 23:10:31  -- 2024 Jun 05 00:36:20   & 5.1    & $C$  \\
VLT/FORS2       & 113.27RE.001    & Wollaston prism + HWP & 2024 Jun 04 23:34:37 -- 2024 Jun 05 00:39:23 & 3.9    &  $R$  \\
\swift/XRT/UVOT  & 00033012241     & PC / Event mode  & 2024 Jun 06 02:39:12 -- 2024 Jun 06 04:48:54  & 2.2  & 0.3--10\,keV / $UVM2$    \\
\nicer/XTI        & 7034060108      &                & 2024 Jun 08 18:53:00 -- 2024 Jun 08 22:17:30   & 1.4  & 0.5--10\,keV    \\
\hline
\end{tabular}
\end{table*}

\subsection{\ixpe\ Observations} 
\ixpe\ observed J1023 during two sessions: May 29--June 2 and June 3--June 14, 2024 (MJD 60459.52--60463.76; MJD 60464.88--60475.62). The total exposure time per detector unit (DU) was $\simeq$675\,ks. To reduce instrumental background, we applied the rejection algorithm by \cite{DiMarco2023}\footnote{\url{https://github.com/aledimarco/IXPE-background}}. Source photons were extracted from a 60\arc-radius circular region centered on the source, and background photons were taken from an annular region with inner and outer radii of 150\arc\ and 300\arc, respectively.
We classified the emission modes based on count-rate thresholds from previous studies  (e.g., \citealt{Bogdanov15, cotizelati18a}) and refined them with a statistical analysis of the \ixpe\ data. A detailed description of the selection process is provided in Appendix\,\ref{sec:ixpe_modes}.
Due to background dominance above 6\,keV, we restricted the polarimetric analysis to the 2--6\,keV band. Data processing was performed using the \texttt{ixpeobssim} package \citep{Baldini2022}. Spectral and spectro-polarimetric analyses were conducted with \texttt{HEASoft} and \texttt{FTOOLS} using the latest calibration files.

\subsection{\nicer\ Observations} 
During the \ixpe\ campaign, we conducted two \nicer\ observations of J1023 (Table\,\ref{tab:log}) to guide the selection of high and low-mode episodes in the \ixpe\ dataset. These observations were performed during orbit night-time to avoid optical light leak due to a damaged thermal shield\footnote{\url{https://heasarc.gsfc.nasa.gov/docs/nicer/analysis_threads/light-leak-overview/}}.
We reprocessed the data using \texttt{NICERDAS} version 13 in \texttt{HEASoft} and calibration products from 2024 February 6. We calibrated and screened the data using \texttt{nicerl2}, and extracted 0.5--10\,keV time series binned at 10\,s using \texttt{nicerl3-lc}. The background level was estimated with the \texttt{SCORPEON} model\footnote{\url{https://heasarc.gsfc.nasa.gov/docs/nicer/analysis_threads/scorpeon-overview/}}.

\subsection{\swift\ Observations} 
We performed two observations with \swift\ simultaneously with the VLA and VLT on 2024 June 4--5 (Table\,\ref{tab:log}), using the X-ray Telescope (XRT) in photon counting mode and the Ultra-Violet/Optical Telescope (UVOT) in event mode with the UVM2 filter (central wavelength 226\,nm; FWHM 52.7\,nm).
Data were processed using calibration files from 2024 May 22 for XRT and 2024 February 1 for UVOT. For XRT, we extracted source photons using a circular aperture of 47.2\arc\ radius and background using an annulus with radii 94.4\arc\ and 188.8\arc. We extracted background-subtracted 0.5--10\,keV time series binned at 50\,s (Fig.\,\ref{fig:timeseries_X}). For UVOT, we used \texttt{coordinator} and \texttt{uvotscreen} to obtain cleaned event lists, and extracted time series binned at 30\,s using \texttt{uvotevtlc} (Fig.\,\ref{fig:timeseries_X}).

\subsection{VLT Observations}
We observed J1023 with the FOcal Reducer/low dispersion Spectrograph 2 (FORS2) on the VLT in polarimetric mode (Table\,\ref{tab:log}). The observations were performed from 2024 June 4 at 23:34:37 UTC to 2024 June 5 at 00:39:23 UTC, under photometric conditions (seeing $\simeq$ 0.3''). A total of 56 images were acquired using the $R_{SPECIAL +76}$ filter ($R$ band; central wavelength 655\,nm; FWHM 165\,nm), each with a 20\,s exposure, totaling 3946\,s. These observations are part of a longer dataset, which includes an additional 3796\,s of observations with the same setup, performed after the conclusion of the observations presented here \citep{Baglio2025}. A Wollaston prism and rotating half-wave plate (HWP) allowed images at four angles $\Phi_i=22.5^{\circ}(i-1)$, $i=1,2,3,4$. Four sets of images were acquired for each angle, repeated 14 times. 

The images were bias-subtracted and flat-fielded. Aperture photometry was performed using the \texttt{daophot} tool \citep{Stetson1987} with a 6-pixel aperture. Using Equations 1--2 of \cite{baglio20}, we calculated the normalized Stokes parameters $Q_{\rm opt}$ and $U_{\rm opt}$. To estimate the degree and angle of linear polarization, we evaluated the parameter $S(\Phi)$ for each of the HWP angles, following \cite{Serego} and references therein (see also \citealt{Covino1999,baglio20, Baglio2023}). This is linked to the  the polarization degree $P_{\rm opt}$ and the angle $\theta$ by the formula: 
\begin{equation}
S(\Phi) = P_{\rm opt}\, \cos [2(\theta-\Phi)].
\end{equation}
We initially estimated the polarization degree $P_{\rm opt}$ and angle $\theta$ by maximizing a Gaussian likelihood function and then refined these estimates using a Markov Chain Monte Carlo (MCMC) procedure \citep{Hogg&Foreman2018} (for further details on the algorithm, see \citealt{baglio20, Baglio2023}). The polarization angle was calibrated using observations of the polarized standard star Vela 1-95, resulting in a correction of $1.6^{\circ}\pm 0.7^{\circ}$.

Instead of relying on a single unpolarized star, we used several field stars (assumed to be intrinsically unpolarized) to correct for both instrumental and interstellar polarization. It is crucial to ensure that the chosen reference stars exhibit similar polarization degrees, which would be a good indicator of their intrinsic non-polarized nature (see e.g., \citealt{baglio20}). Since the field of J1023 is relatively empty of optical sources, we could select only four comparison stars. However, these stars clustered very well together in the $Q_{\rm opt}-U_{\rm opt}$ plane, allowing us to determine the instrumental and interstellar polarization quite precisely in each image. As reported by \cite{Baglio2023}, the maximum interstellar contribution to the linear polarization degree of J1023 is quite low ($P_{\rm opt, int}<0.52\%$), and comparable to the polarization measured for the instrument configuration.

\subsection{VLA Observations} 
\label{sec:vla}
We obtained simultaneous VLA radio observations of J1023 with \ixpe\ and VLT from 2024 June 4 at 23:00:00 UTC to June 5 at 02:59:20 UTC (Project ID 24A-476). The source underwent a bright multi-wavelength flare starting at $\simeq$00:36:00 UTC on June 5. We include here only the pre-flare data, yielding about 1.2\,hr on-source. The full dataset is presented by \cite{Baglio2025}.

We observed in C-band (central frequency $\simeq$6\,GHz) with a 3-bit sampler, dividing the 4096\,MHz band into 32 spectral windows with 64 2-MHz channels each. Calibration used 3C286 (flux, bandpass), J1024$-$0052 (gain), and J1407$+$2827 (leakage). J1023 was monitored in $\simeq$9-min target scans bracketed by two 1-min secondary scans.

Data were processed using the \texttt{CASA} VLA pipeline \citep{Casa2022}, followed by manual and automated (\texttt{rflag} and \texttt{tfcrop}) radio frequency interference flagging. We performed cross-hand calibration following standard VLA polarization procedures\footnote{\url{https://casaguides.nrao.edu/index.php/CASA_Guides:Polarization_Calibration_based_on_CASA_pipeline_standard_reduction:_The_radio_galaxy_3C75-CASA6.5.4}}.
We created Stokes $IQUV$ images with the \texttt{WSClean} imager \citep{WSClean}; \texttt{WSClean} is a CLEAN-based deconvolution algorithm \citep{Hogbom1974} that removes sampling artefacts from interferometric images. To produce our final data products, we first performed a shallow unmasked deconvolution of the full 10$^\prime$ field of view (FOV), i.e., we created preliminary images. We made a deconvolution mask using the unmasked Stokes $I$ image and, with the mask, re-imaged the target field. Adopting the masked image as a model, we performed phase-only self-calibration on the target visibilities with the software package \texttt{QuartiCal} \citep{quartical}. We then refined the mask and repeated the imaging and self-calibration procedure a second time. In the FOV of J1023, there exists, among other fainter background sources, a bright radio Galaxy (J102358.2$+$003826) at an angular offset of ${\sim}\,3^{\prime\prime}$. Using the self-calibrated model, we subtract out the visibilities for the background sources, allowing for narrow FOV images that are minimally affected by aliasing.

We first produced time-resolved $IQUV$ images in 1-min intervals.
These images revealed that J1023 experienced a short ($\sim$8\,min) mini-flare peaking at $\simeq$500\,$\mu$Jy. We then created also four deeper polarization images by combining multiple scans to boost signal-to-noise ratio at intervals corresponding to the times before the mini-flare, the mini-flare, the times after the mini-flare, and the combined times before and after the mini-flare. For each interval, \texttt{WSClean} generated 16 channelized images and a single frequency-averaged Multi-Frequency Synthesis (MFS) image. Flux densities were measured using \texttt{imfit} in \texttt{CASA}, modeling the source as a point source with a synthesized beam-shaped Gaussian fit. $1\sigma$ uncertainties were derived from RMS noise estimates in nearby source-free regions using \texttt{imstat}, employing a circular aperture encompassing $\simeq$$\times$100 synthesized beam areas. For MFS images where J1023 was detected at ${>}20\sigma$ (corresponding to ${>}5\sigma$ in each channelized image), we fitted a power-law of the form $F_\nu \propto \nu^\alpha$ to channelized flux densities to determine the intra-band spectral index $\alpha$.

{Lastly, we produced linear polarization intensity images $(P_{\rm radio} = \sqrt{Q_{\rm radio}^2 +U_{\rm radio}^2})$ for each imaging interval.

\section{Data Analysis and Results}
\label{sec:data_analysis}

Figure\,\ref{fig:lcurves} shows the multiwavelength time series extracted from our dataset, along with the evolution of the optical and radio polarization properties. During the observations, we detected a single episode of low mode at X-ray and UV wavelengths lasting $\approx$5\,min, matching an enhanced flux at both optical and radio wavelengths. In the following sections, we describe the multiwavelength timing and polarimetric properties of J1023 and present the SEDs of the total, pulsed, and polarized emissions.

\begin{figure}
\begin{center}
\includegraphics[width=0.49\textwidth]{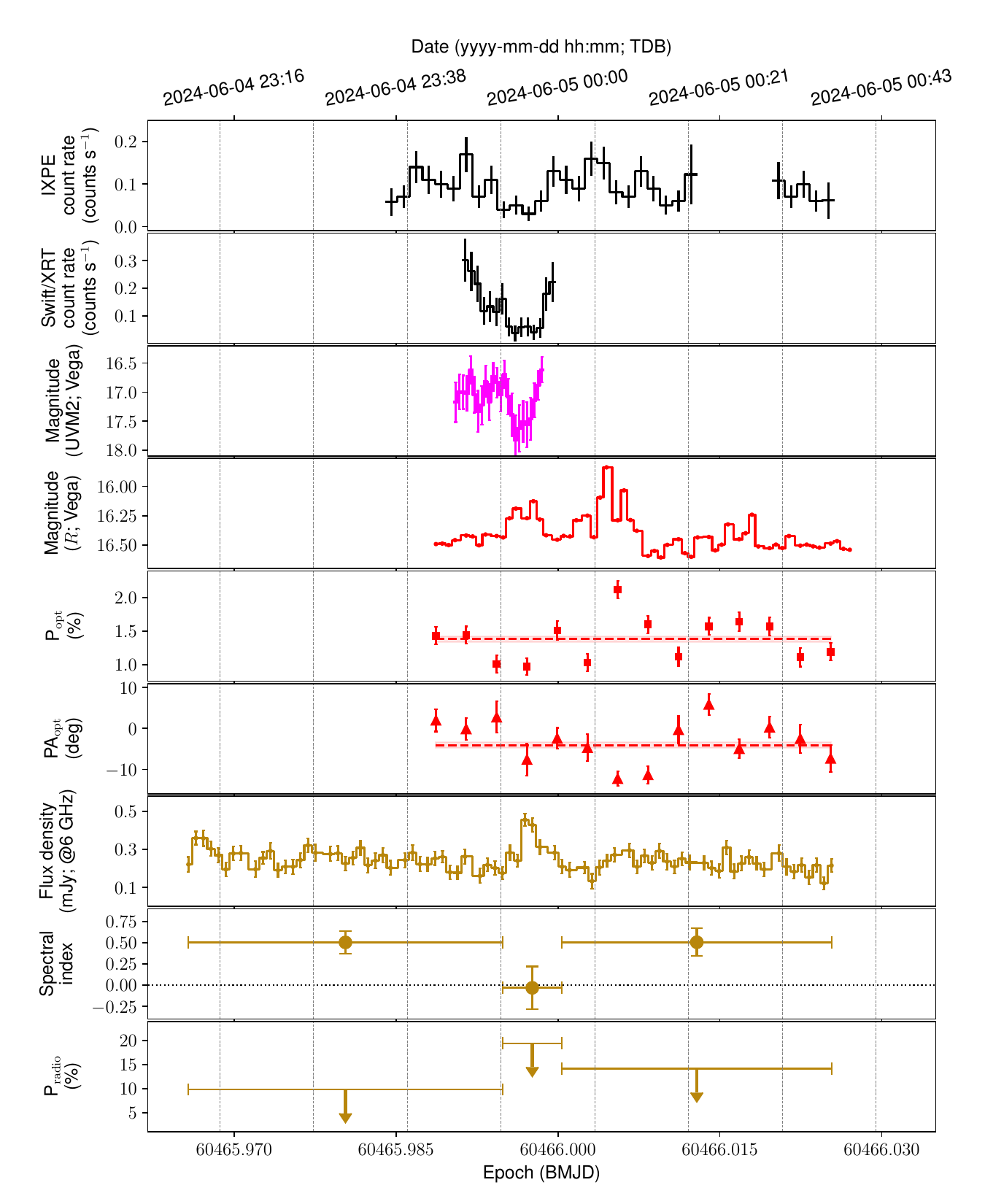}
\vspace{-0.8cm}
\caption{Multiwavelength time series and evolution of optical and radio polarization properties from \ixpe, \swift, VLT, and VLA collected on 2024 June 4--5. Error bars represent 1$\sigma$ c.l. (uncertainties in the fourth panel are smaller than markers). In panels five and six, red horizontal lines show the average optical polarization degree ($P_{\rm opt}$) and angle ($PA_{\rm opt}$), with shaded areas indicating $\pm$1$\sigma$. In the eighth panel, the dotted line marks a flat spectrum. In the ninth panel, downward arrows indicate 3$\sigma$ upper limits on the radio polarization fraction ($P_{\rm radio}$).} 
\label{fig:lcurves}
\end{center}
\end{figure}

\subsection{Timing Analysis of the \ixpe\ Dataset}
Coherent X-ray pulsations at the NS spin period in J1023 are detected only during the high mode \citep{Archibald15, jaodand16, papitto19, illiano23}. Hence, we used source photons collected by \ixpe\ in the high mode in the 2--6\,keV band for our analysis. We corrected photon arrival times for the pulsar's orbital motion, using an orbital period of $P_{\rm orb}$ = 0.1980963155\,d and a projected semi-major axis of asini/c = 0.343356\,light-seconds \citep{illiano23}.
To measure the epoch of passage of the pulsar at the ascending node of the orbit ($T_{\rm asc}$), we performed an epoch folding search around the spin frequency value expected according to the previously measured secular evolution \citep{jaodand16,Burtovoi2020}, sampling each spin cycle in 16 phase bins. We used a grid of $T_{\rm asc}$ values centered on the value predicted according to the long-term evolution measured by \cite{illiano23}, $T_{\rm asc}^{\rm pred}=60461.97948(6)$\,MJD, with a spacing of 0.2\,s. We recovered the coherent signal with an epoch folding variance of $S=103$ (single-trial false alarm rate $p=3.5\times10^{-15}$) at a spin frequency $\nu=592.42146705(9)$\,Hz and $T_{\rm asc}$ = 60461.979580(14)\,MJD. The uncertainty on the spin frequency was evaluated following \cite{Leahy1987}, while that on $T_{\rm asc}$ was taken as the half-width at half-maximum of the Gaussian fit to the pulse variance distribution\footnote{The timing model is archived at Zenodo.}. We obtained compatible results by analysing 0.5--6\,keV \nicer\ data in the high mode taken on 2024 June 1. The timing parameters had slightly larger uncertainties due to the lower number of photons recorded ($\simeq1.9\times10^4$) compared to \ixpe\ ($\simeq2.2\times10^4$) and the uneven coverage. Our measurements tend to favor a scenario where the pulsar's rotational evolution in the subluminous X-ray state remains largely unchanged compared to its behavior during the radio pulsar state (Fig.\,\ref{fig:efsearch}).

\subsection{X-ray Polarimetric Properties}
Details on the polarimetric analysis of the \ixpe\ data are given in Appendix\,\ref{sec:pol_app}. Within the 3--6\,keV band, we detected a polarization degree of $P_X = (16\pm5)$\% and a polarization angle $PA_X = -7\deg \pm 9\deg$, measured counter-clockwise from the North celestial pole toward the East. Hereafter, all uncertainties are reported at 68\% c.l.. This polarization degree exceeds the Minimum Detectable Polarization (MDP; \citealt{Weisskopf2010,Elsner2012}) of 15\% at 99\% confidence, allowing us to reject the null hypothesis of unpolarized emission with 99.4\% confidence (3.2$\sigma$ significance\footnote{\url{https://heasarc.gsfc.nasa.gov/docs/ixpe/analysis/IXPE_Stats-Advice.pdf}}). In the 2--3\,keV band, the polarization degree is consistent with zero within a 1$\sigma$ uncertainty.
A similar analysis of the high-mode data yields values consistent with those reported above, though none exceeded the MDP at 99\% confidence. No significant variations in Stokes parameters were observed across the pulsar rotational cycle or the binary orbital cycle (Appendix\,\ref{sec:ppps_app}). For low-mode and flaring-mode data, the significance was too low for a reliable analysis, yielding upper limits of $P_{\rm X,L} < 26$\% and $P_{\rm X,F} < 28$\%  at 90\% confidence over the 2--6\,keV range (Appendix\,\ref{sec:pol_mi_app}).

To improve the polarimetric measurements, we performed weighted spectro-polarimetric analyses \citep{Strohmayer2017,DiMarco2022} on both the entire dataset and the high-mode dataset (Appendix\,\ref{sec:pol_md_app}). We fitted a power-law model corrected for interstellar absorption \citep{Bogdanov15,cotizelati18a} and convolved with a constant polarization model to the data in the 2--6, 2--3 and 3--6\,keV ranges. Results are summarized in Table\,\ref{tab:X_pol}. For the average emission in the 2--6\,keV range, we obtained $P_X = (11\pm3)$\% at $PA_X = -7\deg\pm8\deg$. For the high-mode emission in the same range, we measured $P_{\rm X,H} = (12\pm3)$\% at $PA_{\rm X,H} = -2\deg\pm9\deg$. To estimate the significance level, we used the \texttt{steppar} command in \texttt{xspec} on PD and PA. The contour remains open at a significance above 99.7\%. Figure\,\ref{fig:polar_plot} shows the confidence contours for the measurements of $P_{\rm X,H}$ and $PA_{\rm X,H}$ in the high mode across energy bands.

Previous analyses show that around 10\% of low-mode episodes in the \ixpe\ datasets may be missed due to their short durations \citep{Archibald15,Bogdanov15,cotizelati18a}. However, this does not affect the polarization degree measured during the high mode, as the potential increase, about 0.2\% in the polarization degree, is within the measurement uncertainties (see Appendix\,\ref{sec:pol_md_app}).

\begin{table}
\centering
\footnotesize  
\setlength{\tabcolsep}{2pt} 
\renewcommand{\arraystretch}{1.2}

\caption{Polarization degree ($P_X$), angle ($PA_X$), and fit statistics ($\chi^2$, dof) measured in distinct X-ray energy bands by combining data from the three \ixpe\ DUs using spectro-polarimetric analysis. Uncertainties are quoted at the 68.27\% c.l.}
\label{tab:X_pol}

\resizebox{\columnwidth}{!}{
  \begin{tabular}{c c c c | c c c}
  \hline\hline
  Energy Band (keV) & \multicolumn{3}{c|}{Average Emission} & \multicolumn{3}{c}{High Mode} \\
  \cline{2-7}
  & $P_{\rm X}$ (\%) & $PA_{\rm X}$ (\deg) & $\chi^2$/dof & $P_{\rm X,H}$ (\%) & $PA_{\rm X,H}$ (\deg) & $\chi^2$/dof  \\
  \hline
  2--6  & $11 \pm 3$  & $-7 \pm 8$  & 175/166  & $12 \pm 3$  & $-2 \pm 9$  & 164/166 \\
  \hline
  2--3  & $8 \pm 5$   & $20 \pm 20$ &  32/31   & $8 \pm 6$   & $10 \pm 20$ &  25/31  \\
  \hline
  3--6  & $15 \pm 4$  & $-9 \pm 8$  & 120/121  & $15 \pm 5$  & $-3 \pm 9$  & 120/121 \\
  \hline
  \end{tabular}%
}
\end{table}

\begin{figure}
\begin{center}
\includegraphics[width=0.45\textwidth]{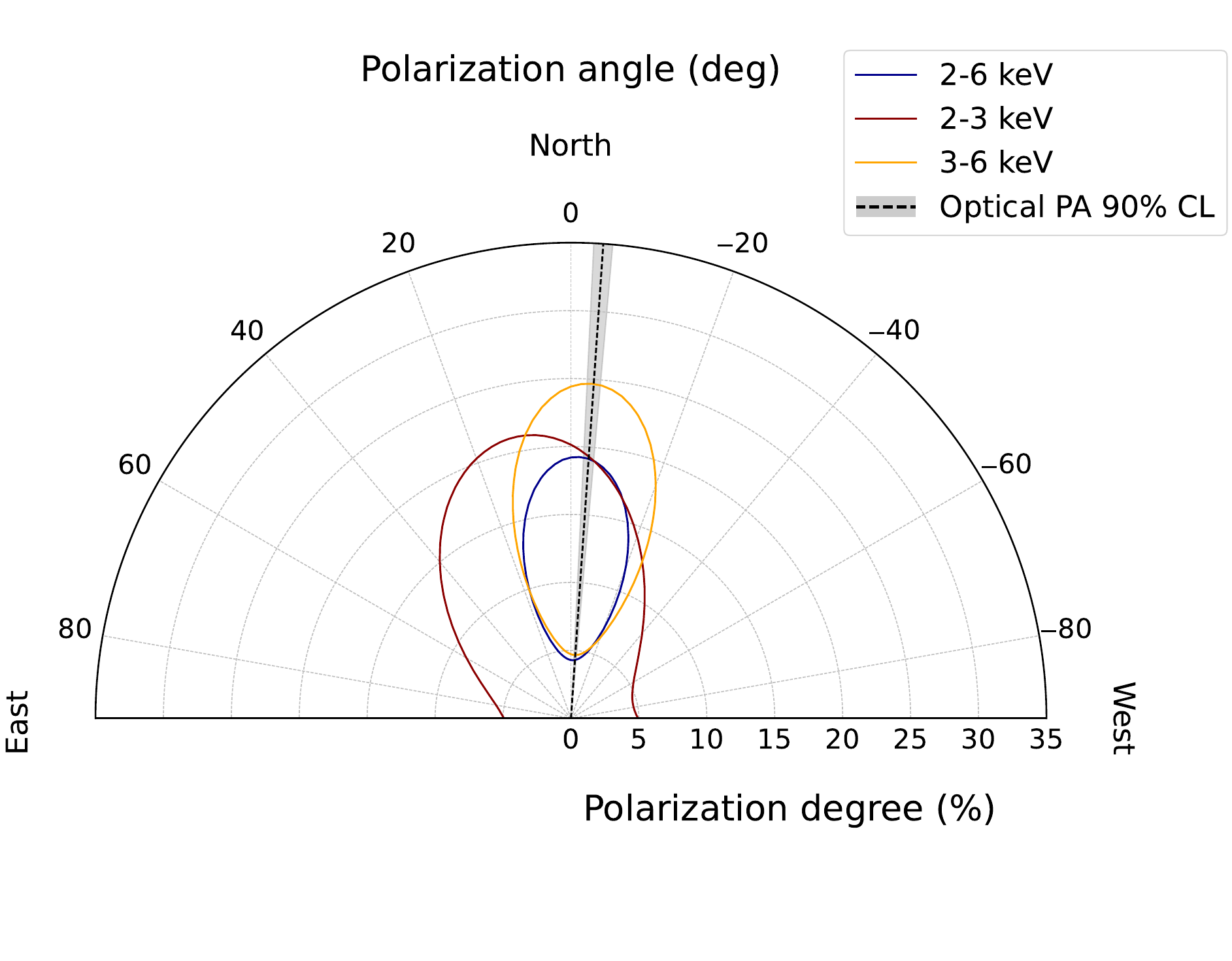}
\vspace{-1cm}
\caption{Protractor plot of X-ray polarization degree and angle for the high mode, derived from a weighted spectro-polarimetric analysis, compared with the optical polarization angle. The polarization degree and angle are measured over the energy ranges 2--6, 2--3, and 3--6\,keV, and are displayed along the radial and azimuthal directions, respectively. Contours represent 90\% confidence regions. The black dashed line and gray shaded area indicate the optical $R$-band polarization angle and its 90\% c.l. uncertainty.} 
\label{fig:polar_plot}
\end{center}
\end{figure}

\subsection{Optical Properties}
Details on the optical properties of J1023 are presented in Appendix\,\ref{sec:opt_app}.
J1023 was detected at an average magnitude of $\simeq$16.4 (Vega) in the $R$-band.
We measured an average linear polarization degree of $P_{\rm opt} = (1.38\pm 0.04)\%$ and a polarization angle of $PA_{\rm opt} = -4.1\deg\pm0.7\deg$.  
During the X-ray high mode, we measured $P_{\rm opt,H} = (1.41 \pm 0.04)\%$ at an angle $PA_{\rm opt,H} = -3.9^\circ \pm 0.7^\circ$. This polarization angle is fully consistent with that derived at X-ray energies within the uncertainties (Fig.\,\ref{fig:polar_plot}). During the X-ray low mode, the polarization degree decreases slightly to $P_{\rm opt,L}=(0.97\pm 0.13)\%$. However, the polarization angles in high and low modes remain statistically consistent with each other.

\subsection{Radio Properties}
\label{sec:data_analysis_radio}
J1023 exhibited an average flux density of $\simeq$240\,$\mu$Jy at 6\,GHz and experienced a mini-flare matching a low-mode episode in X-rays lasting $\simeq$8\,min, reaching a peak flux density of $\simeq$500\,$\mu$Jy (Fig. \ref{fig:radio_properties}). The power-law  spectral index $\alpha$ remains nearly constant during the high-mode episodes that precede (H1) and follow (H2) the low mode, but decreases during the low mode (L): we obtain $\alpha_{\rm H1} = 0.50 \pm 0.13$, $\alpha_{\rm H2} = 0.51 \pm 0.16$ and $\alpha_{\rm L} = -0.03 \pm 0.25$.  These results are consistent with previous observations \citep{bogdanov18} which pointed to optically thick synchrotron emission in the high mode and a mix of optically thick and optically thin synchrotron emission in the low mode \citep{Baglio2023}.

\begin{figure}
\begin{center}
\includegraphics[width=0.48\textwidth]{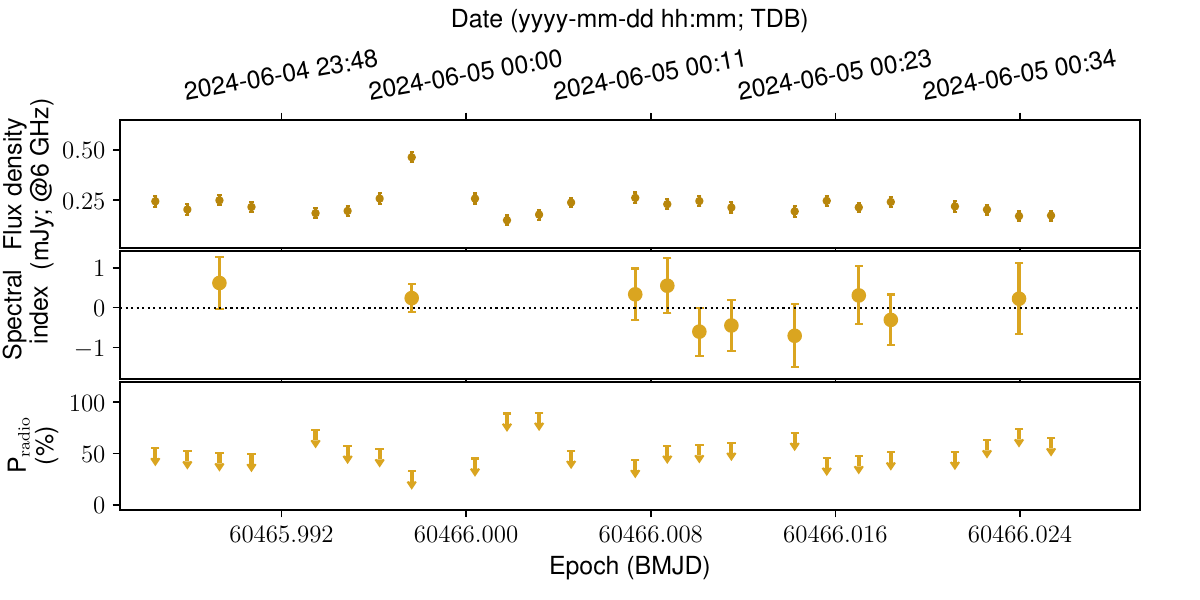}
\vspace{-0.5cm}
\caption{Evolution of the radio properties. The panels, from top to bottom, illustrate the temporal evolution of the radio flux density at the central frequency of 6\,GHz in the VLA's C-band; the spectral index, calculated by dividing the 4–8\,GHz band into two sub-bands (4–6\,GHz and 6–8\,GHz); and the 3$\sigma$ upper limits on the linear polarization fraction. Each time bin represents a 2-min interval. Spectral indices are only shown when they are constrained, with the horizontal dotted line marking a spectral index of 0 (i.e., a flat spectrum). BMJD stands for Barycentric Modified Julian Date, TDB for Barycentric Dynamical Time.}
\label{fig:radio_properties}
\end{center}
\end{figure}

No radio polarization was detected (Fig. \ref{fig:radio_properties}). Therefore, we derived 3$\sigma$ ($\simeq$\,99.7\%) upper limits on the polarization fraction from the $P$-images following the prescription outlined by \cite{Vaillancourt2006}. We established $P_{\rm radio,H1} <10\%$, $P_{\rm radio,H2} <14\%$ and $P_{\rm radio,L} <19\%$. By stacking high-mode data, we achieved $P_{\rm radio,H} <9\%$. These results represent the first constraints on the linear polarization fraction of the radio emission in both modes.

\subsection{Spectral Energy Distributions}
We extracted the SEDs for the high mode (see Fig.\,\ref{fig:SEDs}) by combining the unabsorbed X‑ray fluxes in the 2--3, 3--4 and 4--6\,keV energy bands from \ixpe\ with the averaged $R$-band fluxes from VLT/FORS2. Adopting the estimated hydrogen column density ($N_H = 2.8 \times 10^{20}\, \rm cm^{-2}$), we calculated the $V$-band absorption coefficient based on the relation from \cite{Foight2016} and determined the $R$-band absorption using the extinction laws from \cite{Gordon2023} (see also \citealt{Gordon2009,Fitzpatrick2019,Gordon2021,Decleir2022}). 

For the polarized SED, we used the polarization degrees measured in the 2--3\,keV and 3--6\,keV bands through spectro-polarimetric analyses (see Table\,\ref{tab:X_pol}). The dereddened polarized optical flux was estimated by rescaling according to the polarization degree in the high mode, $(1.41 \pm 0.04)\%$.

For comparison, we overlaid the total and pulsed X-ray and optical fluxes obtained from a previous campaign involving the \xmm\ and \nustar\ satellites as well as the SiFAP2 photometer mounted on the INAF \emph{Telescopio Nazionale Galileo} \citep{papitto19}. Additionally, we included the fit of their SED of the pulsed emission using a single power law defined by $F_{\nu} = a \cdot \nu^{b}$, where \( F_{\nu} \) is the flux density at frequency \( \nu \), \( a \) is the normalization constant, and \( b \) is the spectral index. Using nonlinear least squares fitting, the best-fit parameters were determined to be \( a = (2.86 \pm 0.01) \times 10^{-13}\, \mathrm{erg\,cm^{-2}\,s^{-1}} \), and \( b = 0.288 \pm 0.015 \) \citep{papitto19}.

Our analysis shows that the SED of the polarized flux is consistent with the same power-law model that describes the spectrum of the pulsed signal. This provides additional strong evidence that the observed polarized emission shares a common origin with the pulsed emission across both optical and X-ray bands. Moreover, the near-coincidence of the polarized and pulsed fluxes throughout the spectrum indicates that the observed polarization originates almost entirely from the same emission processes responsible for the pulsations across the optical and X-ray frequencies. Theoretical predictions for the linear polarization of synchrotron radiation, assuming a spectral index of $\alpha \simeq1.7$ (as measured by \citealt{papitto19} and shown in Fig. \ref{fig:SEDs}), suggest a theoretical maximum polarization degree of $\simeq$77\% \citep{Rybicki}, which is consistent with our findings. Future instrumentation designed to measure the polarization of pulsed emission will be essential to test this.

 \begin{figure}
    \centering
    \includegraphics[width=1.0\linewidth]{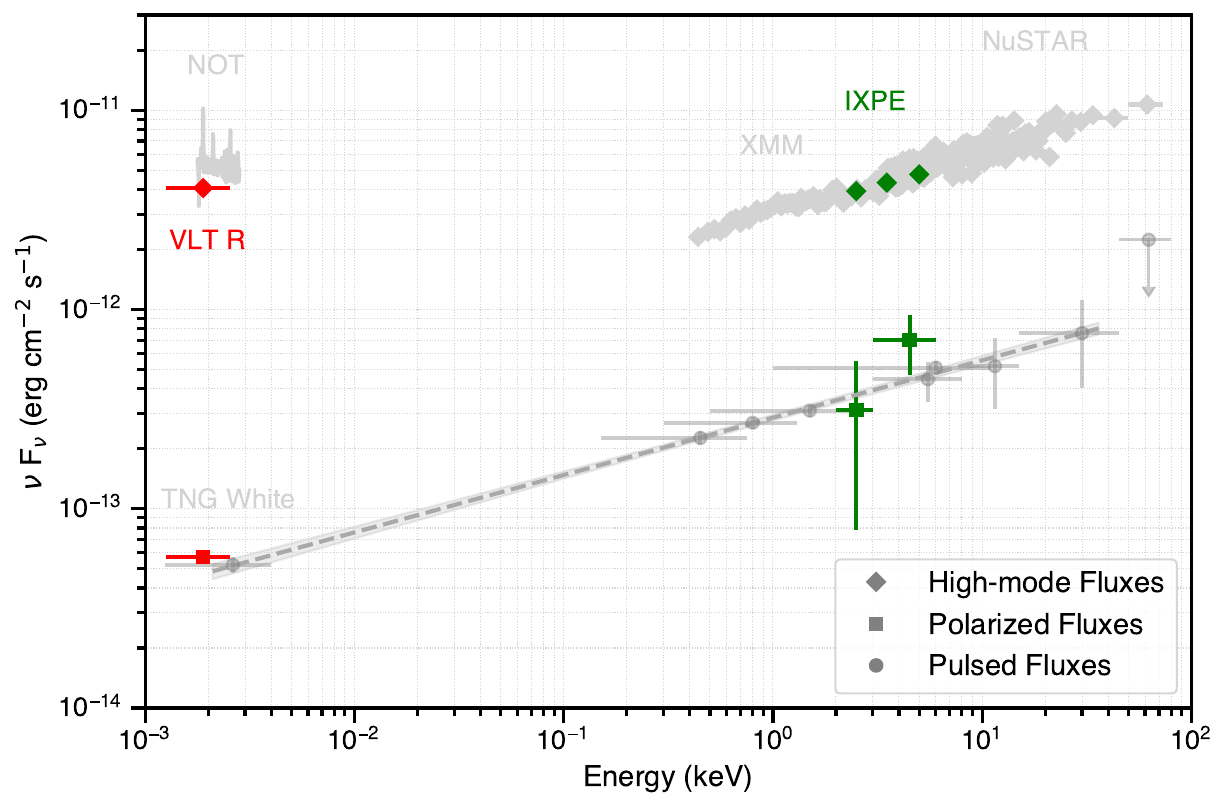}
    \vspace{-0.7cm}
    \caption{SEDs for the high mode of J1023 derived from \ixpe\ and VLT high-mode fluxes (green and red diamonds), with polarized fluxes depicted as squares. High-mode fluxes and pulsed fluxes in the X-ray and optical bands derived by \cite{papitto19} are shown as light-gray diamonds and circles, respectively. The gray dashed line represents the power-law fit to the pulsed fluxes.}
    \label{fig:SEDs}
\end{figure}

\section{Discussion}
\label{sec:discussion}

\subsection{The Nature of the Polarized Emission}
\subsubsection{An Accretion-powered Mechanism?}

Theoretical models for accreting NSs in LMXBs predict that the primary X-ray emission mechanisms (including thermal Comptonization, emission from the spreading/boundary layer, disk reflection, and fan-beamed radiation from small, heated surface spots) should yield polarization degrees of only a few percent (for a review, see \citealt{Ursini2024}). For example, standard Comptonization models predict that seed photons scattering off high‑energy electrons yield only modest polarization. Similarly, theoretical studies constrain the polarization from emission in the spreading layer to be $\lesssim$1.5\% \citep{Bobrikova2023,Bobrikova2024,Farinelli2024}. These theoretical expectations are well supported by the \ixpe\ observations conducted so far on high-luminosity accreting NSs (with X-ray luminosities at least three orders of magnitude higher than J1023), where these processes resulted in polarization degrees of only a few percent (see \citealt{Papitto2025} for the case of an accreting MSP).

In stark contrast, our observations of J1023 show a substantially higher polarization degree in the X-ray band. Furthermore, the pure power-law spectrum of J1023 lacks additional components (e.g., multiple Comptonized components, disk reflection features, or signatures of localized fan-beamed emission) that would be expected if the above-mentioned standard mechanisms were significantly contributing to the X-ray emission. This discrepancy strongly suggests that an additional or alternative mechanism must be at work to generate the much higher polarization degree observed from J1023.

Scattering of emission by outflows of highly ionized material launched from the accretion disk along the equatorial plane -- commonly referred to as disk winds (e.g. \citealt{DiazTrigo2016}) -- may also provide a source of polarization. Although disk winds have not been observed in J1023, recent work \citep{Nitindala2025} suggests that even if they were present and contributed significantly to the X‑ray emission, such winds alone could not account for the high polarization degree given the system’s orbital inclination of $i = 46.4^{+0.5}_{-0.7}\deg$; \citealt{Stringer2021}).

\subsubsection{A Rotation-powered Mechanism?}
A few models have been developed to explain the X-ray to optical polarization properties in isolated pulsars, including the outer gap and two-pole caustic models, as well as current sheet models (for a review, see \citealt{Harding2019}). In what follows, we focus on our \ixpe\ results in the 2--6\,keV band (see Fig.\,\ref{fig:pol_phases}) and compare them with the predictions from these models, keeping in mind that the photon statistics in our dataset limit the degree to which we can definitively confirm or rule out any particular scenario.

The outer gap (OG; \citealt{Cheng1986,Cheng2000,Harding2001,Takata2008}) and two-pole caustic (TPC; \citealt{Dyks2003}) models propose that the high-energy emission of pulsars originates from the outer magnetosphere, along the last open magnetic field lines inside the light cylinder radius. The emission mechanisms involve synchrotron radiation and curvature radiation from particles accelerated within vacuum gaps or extended regions of strong electric fields. Pulse profiles are characterized by caustics formed due to relativistic effects like aberration and light travel time delays. This results in double-peaked light curves, with peak separation depending on the magnetic inclination angle and the observer's viewing angle. These models predict polarization properties such as rapid swings in the PA correlated with the pulses and dips in PD at the pulse peaks. The dips are attributed to depolarization from overlapping emissions originating from regions with varying magnetic field orientations. In the TPC model, rapid PA swings are expected through both peaks of double-peaked pulse profiles. In contrast, the OG model predicts a significant PA swing primarily at the second peak, and the first peak may not show a strong PA swing unless the emission occurs very near the light cylinder \citep{Dyks2004}.

The X-ray PA in J1023 remains relatively constant across the pulse profile within the uncertainties, showing variations by a few tens of degrees at most (Fig. \ref{fig:pol_phases}). This is at odds with the strong, rapid PA swings of up to 180\deg\ predicted by OG/TPC models \citep{Harding2019}.

Current sheet models \citep{Petri2005,Cerutti2016} focus on emission from particles accelerated in the equatorial current sheet beyond the light cylinder. Magnetic reconnection in the current sheet accelerates particles, leading to high-energy synchrotron emission. Pulses are formed when the observer's line of sight crosses the current sheet, resulting in one or two pulses per rotation. These models predict PD values typically $\simeq$15–30\%, with PD dips at the pulse peaks due to depolarization effects. Significant PA swings are expected at the pulse peaks due to the reversal of the magnetic field polarity across the current sheet.

J1023 does not exhibit the sharp X-ray PA swings predicted by current sheet models (Fig. \ref{fig:pol_phases}). This lack of pronounced PA variation suggests a more uniform magnetic-field orientation in the emission region, inconsistent with the magnetic field reversals expected in current sheet models. 

Overall, we find no clear evidence for the pronounced polarimetric swings predicted by standard rotation-powered pulsar emission models. Although the limited photon statistics in some phase bins prevent us from excluding milder geometric variants of these models, the absence of prominent polarimetric swings at pulse maxima in the X-ray band is more consistent with a relatively uniform magnetic-field configuration rather than with the dramatic field reversals typically required by such models.
 
\subsubsection{Pulsar Wind -- Accretion Flow Interaction?}
\label{sec:mini-pwn}
A scenario for the high mode of J1023 suggests the presence of an active rotation-powered pulsar, an accretion disk, and a compact jet. In this scenario, the observed X-ray, UV, and optical pulses originate from synchrotron emission at a boundary region where the pulsar electromagnetic wind -- modulated at the NS spin period -- collides with the inner accretion flow \citep{papitto19,Veledina19}. This interaction is thought to occur just beyond the light cylinder radius (the distance at which corotation with the pulsar becomes physically impossible), which is $\simeq$80\,km for J1023. The observed lag of $\simeq$150\,$\mu$s between optical and X-ray pulses is suggested to originate from differences in the cooling timescales of synchrotron photons at these energies, given the expected magnetic field strength at the boundary region \citep{papitto19, illiano23}. A compact jet is launched likely near the compact object, emitting partially self-absorbed synchrotron radiation from radio to infrared frequencies. Previous work shows that the emission at the boundary region contributes $\simeq$3\% of the optical flux and $\simeq$83\% of the 0.3--10\,keV flux, while the compact jet emission accounts for $\simeq$2.5$\%$ in the optical band and $\simeq$17\% in the 0.3--10\,keV energy band \citep{Baglio2023}.

Electrons spiraling along toroidal magnetic field lines around the jet axis emit synchrotron radiation, with the polarization oriented perpendicular to the field lines. 
Given this magnetic field geometry and the orbital inclination of J1023, calculations indicate that the expected polarization degree should not exceed $\simeq$25\% \citep{Veledina_Pelissier2025,Lyutikov2005,Krawczynski2022}. In the $R$-band, this translates to a maximum polarization degree of $\simeq$0.6\%, which is lower than the average optical polarization observed for J1023. In the \ixpe\ energy band, the limit is $\simeq$3\%, again below our measured value. Hence, based on these considerations, we rule out the compact jet as the main source of the observed polarization.

If the observed low-level optical polarization is due to synchrotron radiation from the boundary region, and a single population of electrons accelerated at this region produces synchrotron radiation from optical to X-rays, then the intrinsic polarization in the optical band should extend to higher energies. In this scenario, we estimate an intrinsic polarization degree of 12--17\% in the X-ray band and expect close alignment between the average optical and X‑ray polarization angles \citep{Baglio2023}. Our measurements are fully consistent with these predictions, providing striking evidence that the polarized emission in J1023 is driven by emission at the boundary region. 

Figure\,\ref{fig:SEDs} shows the SEDs of the high-mode flux, polarized flux, and pulsed flux of J1023 across optical and X-ray wavelengths. The SED of the polarized flux follows the same power law that describes the SED of the pulsed flux. Moreover, both the polarized and pulsed fluxes are consistent with each other. These results suggest that the polarized and pulsed emissions share a common origin, lending support to the scenario in which the synchrotron-emitting boundary region is the primary source of both in J1023.

\begin{figure}
\begin{center}
\includegraphics[width=0.45\textwidth]{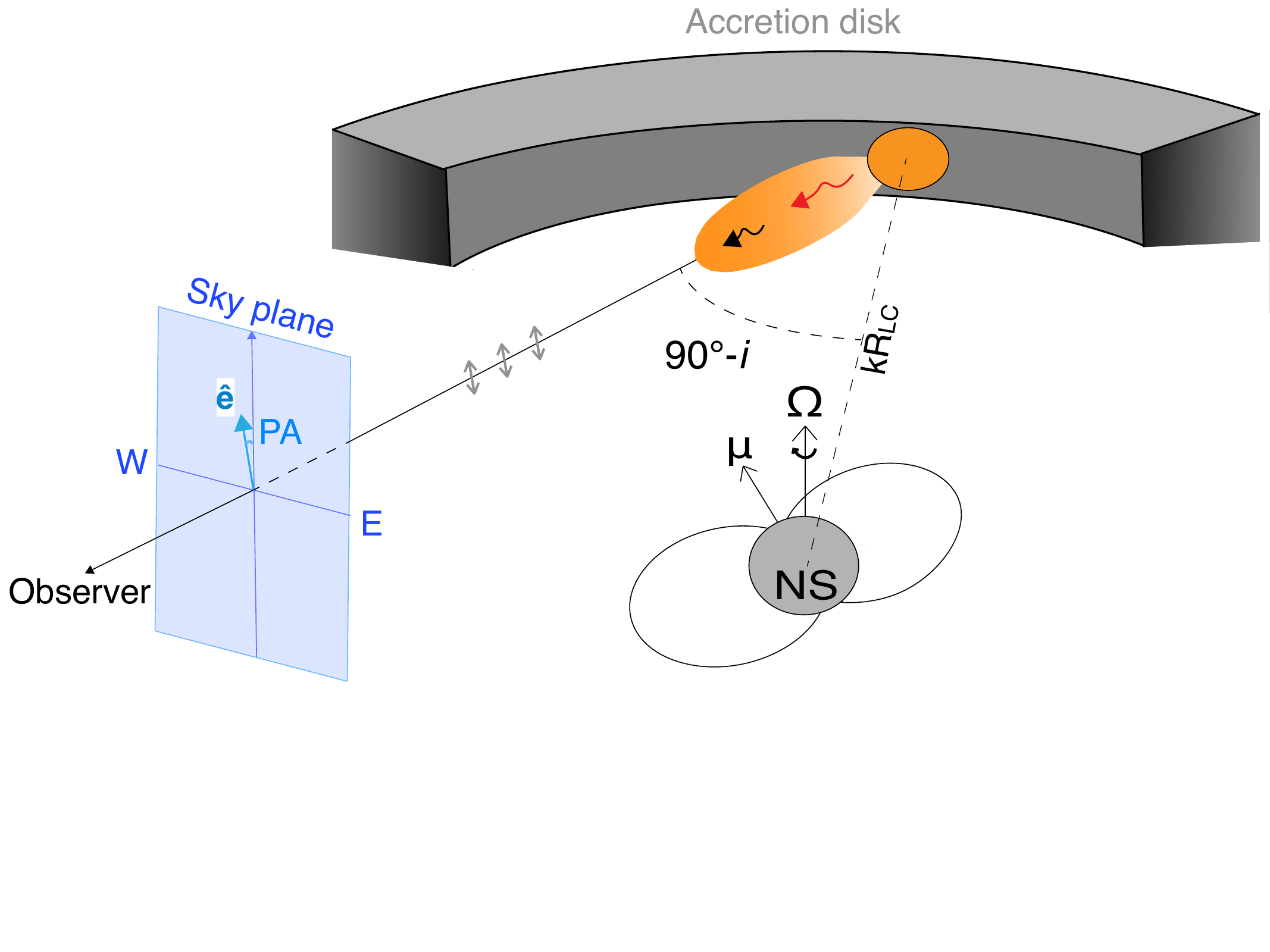}
\vspace{-1.7cm}
\caption{Geometry of polarized emission in the high mode of J1023. Pulsed emission originates from synchrotron emission at the boundary region where the pulsar electromagnetic wind -- modulated at the NS rotational period -- collides with the inner accretion flow at a distance of $kR_{\rm LC}$ ($k \simeq 1$--2; $R_{\rm LC}$ is the light cylinder radius; \citealt{papitto19,Veledina19}). The magnetic field at the boundary region comprises both toroidal and poloidal components, with the constant X-ray $PA$ across the pulsar phase indicating that the poloidal component is likely dominant. Particles in the region visible to the observer (orange filled circle) emit synchrotron radiation boosted perpendicular to the surface of the boundary region (orange shaded area). Black and red wave-like arrows represent X-ray and optical photons emitted from the boundary region. The average $PA$ is defined as the angle between the North-South direction on the sky plane (cyan parallelogram) and the projected electric field vector (${\bf \hat{e}}$). Gray double-ended arrows represent the amplitude of the electric field (not to scale). \textbf{$\mu$} and \textbf{$\Omega$} denote the pulsar magnetic moment and angular velocity vector, respectively.} 
\label{fig:sketch}
\end{center}
\end{figure}


Figure\,\ref{fig:sketch} illustrates the geometry of the system in the high mode. 
The $PA$ of synchrotron radiation traces the orientation of the magnetic field projected onto the sky plane, perpendicular to the observer’s line of sight. If the observed pulsed emission from optical to X-ray energies arises from synchrotron radiation emitted by particles accelerated at regions on the NS wind -- disk boundary that corotate with the NS \citep{papitto19}, the magnetic field geometry strongly influences how the $PA$ varies with pulsar rotation. A predominantly toroidal magnetic field at the emission site would produce a periodic modulation of the $PA$ by 180\deg\ each rotation. Conversely, a predominantly poloidal magnetic field configuration would yield a nearly constant, phase-independent $PA$ \citep{Veledina_Pelissier2025}. Although a modest toroidal component cannot be entirely ruled out, given statistically significant $PA$ variations observed in a few pulse-phase bins, the overall consistency of the $PA$ across the spin cycle (Fig.\,\ref{fig:pol_phases}) suggests a predominantly poloidal field geometry. Because the observed synchrotron emission integrates contributions from many small regions across the wind--disk boundary -- each characterized by similar magnetic-field orientations and electron energy distributions -- local fluctuations average out, producing a $PD$ that remains nearly constant throughout the pulsar rotation.

\subsection{Synchrotron Emission across Scales}
Our findings reveal striking similarities between J1023 and pulsar wind nebulae (PWNe). In PWNe, the high-energy emission originates from synchrotron radiation by electrons accelerated within highly ordered magnetic fields. These fields are shaped by the pulsar wind as it wraps around the pulsar and is compressed within the nebula (see, e.g., \citealt{Reynolds2012}).
However, PDs in PWNe typically exceed those measured in J1023. For example, the Vela PWN exhibits a X-ray PD averaging $\sim$45\%, reaching up to 70\% in inner regions \citep{Xie2022}. Additionally, a PD exceeding 60\% has been detected in the outskirts \citep{Liu2023}. The Crab PWN shows a spatially integrated PD of $\approx$20\%  \citep{Bucciantini2023}, with some regions exhibiting PDs of 45 -- 50\% \citep{Liu2023,Bucciantini2023,Zuo2025}. The PWN around the pulsar PSR\,B0540$-$69 reaches a PD of $\simeq$25\% \citep{Xie2024}. The PWN 3C\,58 is characterized by an integrated PD of $\simeq$22\% \citep{Bucciantini2025}.

The distinct geometries and spatial scales of PWNe compared to J1023 may play a crucial role in the observed differences. In PWNe, the pulsar wind expands into the surrounding environment over large distances, with the wind termination shock occurring at $\approx$0.1\,pc from the pulsar \citep{Hester2002,Petre2007}. This extended region facilitates the development of large-scale, toroidal magnetic fields with highly ordered structures. At the same time, it disperses any coherent pulsations emitted by the NS. Consequently, PWNe are not expected to produce pulsed emission through the same emission mechanism described for J1023. In contrast, J1023 features a much more compact emission region located $\approx$100\,km from the pulsar. At such close proximity, the interaction between the pulsar wind and the inflowing material from the accretion disk is more disruptive to the striped wind structure due to the higher density and turbulence of the accretion flow. This interaction likely results in a more tangled magnetic field at the boundary region, which may be the main factor accounting for the lower polarization degree observed in J1023 compared to PWNe.

An especially relevant comparison comes from recent \ixpe\ observations of the binary pulsar PSR\,B1259$-$63 \citep{Kaaret2024}, where the pulsar wind interacts with a massive companion's decretion disk. Despite the different source of the interacting matter, PSR\,B1259$-$63 shows a PD ($\simeq$8\%) similar to J1023. This similarity reflects a common scenario in binary systems, where the interactions between pulsar winds and dense matter flows in compact and turbulent boundary regions inherently produce less ordered magnetic fields and thus reduced polarization in the emitted radiation, compared to the highly structured fields seen in PWNe.

\section{Conclusions}
\label{sec:conclusions}

We have conducted the first multiwavelength polarimetric campaign of the tMSP PSR\,J1023$+$0038, using observations across the X-ray, optical, and radio bands. The results of our analysis can be summarized as follows:

$\bullet$ Timing analysis of the \ixpe\ data revealed coherent X-ray pulsations at the NS spin period. The measured spin frequency suggests that the rotational evolution of the pulsar in its subluminous X-ray state has remained largely unchanged compared to its behavior during the radio pulsar state.

$\bullet$ During the high mode, we detected polarized X-ray emission with a polarization degree of $(12 \pm 3)\%$ and angle of $-2^\circ \pm 6^\circ$ (1$\sigma$) in the 2--6\,keV energy range. During the low mode, we derived an upper limit on the polarization degree of 26\% at 90\% c.l.. We do not observe any significant variability in the polarization properties along the NS rotational phases or the binary orbital phases.

$\bullet$ Optical polarization measurements revealed a polarization degree of $(1.41 \pm 0.04)\%$, with a polarization angle aligning closely with that of the X-ray emission. This strongly suggests a common origin for the polarized emission in both the optical and X-ray bands.

$\bullet$ No polarized emission was detected at radio wavelengths, down to 3$\sigma$ upper limits on the polarization fraction of 9\% in the high mode and 19\% in the low mode. 

$\bullet$ The SEDs of the polarized and pulsed fluxes align with a single power-law model across optical and X-ray energies. This suggests that the polarized emission originates from the same processes responsible for the pulsed emission.

$\bullet$ The polarization degree in the high mode is higher than that seen in accreting NSs in LMXBs, indicating that standard accretion-powered emission mechanisms cannot fully explain the observed polarization. Additionally, scattering in an accretion disk wind alone is unlikely to account for the high polarization degree. Moreover, the lack of significant variability in the X-ray polarization properties across the pulsar rotational phases disfavors scenarios where the polarized signal arises as if the NS behaves like an isolated pulsar.

$\bullet$ Our findings support a scenario where the polarized emission is produced by synchrotron radiation at a boundary region formed by the interaction between the pulsar wind and the inner accretion disk. This gives insights into how the pulsar wind interaction with its environment shapes the observed emission properties.

These findings enhance our understanding of the complex interplay between accretion-powered and rotation-powered processes in tMSPs, and establish multiwavelength polarimetry as a crucial diagnostic tool for uncovering the physical mechanisms at work in these systems. Future observations, particularly those capable of measuring the polarization of the multiwavelength pulsed emission, will be essential to further elucidate the nature of the emission processes in J1023 and similar objects.

\begin{acknowledgments}
We thank the referee for their valuable suggestions, which have helped improve the clarity and quality of the manuscript.
We thank the \nicer\ principal investigator, Keith Gendreau, for approving our Target of Opportunity (ToO) request and the operation team for executing the observations; we also thank the \swift\ deputy project scientist, Brad Cenko, and the \swift\ duty scientists and science planners, for making the \swift\ ToO observations possible. We thank Alexandra Veledina, Juri Poutanen and Beno\^it Cerutti for their insightful discussions that improved our interpretation of the polarization data analysis results. We thank Alessio Marino for useful discussions on the spectral properties of low-mass X-ray binaries.

This research used data products provided by the \ixpe\ Team (MSFC, SSDC, INAF, and INFN) and distributed with additional software tools by the High-Energy Astrophysics Science Archive Research Center (HEASARC), at NASA Goddard Space Flight Center (GSFC). This research is based on observations collected at the European Southern Observatory under ESO programme 113.27RE. \ixpe\ is a joint US and Italian mission. \nicer\ is a 0.2--12\,keV X-ray telescope operating on the International Space Station, funded by NASA. The National Radio Astronomy Observatory is a facility of the National Science Foundation operated under cooperative agreement by Associated Universities, Inc.  

M.C.B. acknowledges support from the INAF-Astrofit fellowship.
F.C.Z. acknowledges support from a Ram\'on y Cajal fellowship (grant agreement RYC2021-030888-I).
S.C. and P.D.'A. acknowledge support from ASI grant I/004/11/5.
A.D.M. and F.L.M. are supported by the Italian Space Agency (Agenzia Spaziale Italiana, ASI) through contract ASI-INAF-2022-19-HH.0 and by the Istituto Nazionale di Astrofisica (INAF) in Italy.
A.P. and D.d.M. acknowledge financial support from the Italian Space Agency (ASI) and National Institute for Astrophysics (INAF) under agreements ASI-INAF I/037/12/0 and ASI-INAF n.2017- 14-H.0 and from INAF `Sostegno alla ricerca scientifica main streams dell`INAF', Presidential Decree 43/2018, from INAF `SKA/CTA projects', Presidential Decree 70/2016. 
F.C.Z., S.Ca., P.D.'A., A.P., D.d.M. and G.I. acknowledge financial support from INAF-Fundamental research astrophysics project ``Uncovering the optical beat of the fastest magnetised neutron stars'' (FANS)  and the Italian Ministry of University and Research (MUR) under PRIN 2020 grant No. 2020BRP57Z ``Gravitational and Electromagnetic-wave Sources in the Universe with current and next-generation detectors (GEMS)''.
AP acknowledges support from the Fondazione Cariplo/Cassa Depositi e Prestiti, grant no. 2023-2560.
A.K.H. is supported by NSERC Discovery Grant RGPIN-2021-0400.
D.M.R. and K.A. are supported by Tamkeen under the NYU Abu Dhabi Research Institute grant CASS.
D.F.T. is supported by the grant PID2021-124581OB-I00 funded by MCIU/AEI/10.13039/501100011033 and 2021SGR00426.
F.C. acknowledges support from the Royal Society through the Newton International Fellowship programme (NIF/R1/211296).
G.I. is supported by the AASS Ph.D. joint research program between the University of Rome ``Sapienza'' and the University of Rome ``Tor Vergata'' with the collaboration of the National Institute of Astrophysics (INAF).
N.R. is supported by the European Research Council (ERC) under the European Union's Horizon 2020 research and innovation programme (ERC Consolidator Grant ``MAGNESIA'' No. 817661) and the Proof of Concept ``DeepSpacePulse'' (No. 101189496). F.C.Z. and N.R. acknowledge support from grant SGR2021-01269 from the Catalan Government.
This work was also supported by the Spanish program Unidad de Excelencia Mar\'ia de Maeztu CEX2020-001058-M and by MCIU with funding from European Union NextGeneration EU (PRTR-C17.I1). 

The data that support the findings of this study are publicly available on their respective online archive repositories (\ixpe: \url{https://heasarc.gsfc.nasa.gov/FTP/ixpe/data/obs/03/03005599/}; \nicer: \url{https://heasarc.gsfc.nasa.gov/docs/nicer/nicer_archive.html} under ObsIDs 7034060107 and 7034060108; \swift: \url{https://heasarc.gsfc.nasa.gov/cgi-bin/W3Browse/swift.pl} under ObsIDs 00033012239 and 00033012240; VLT/FORS2: \url{http://archive.eso.org/}; VLA: \url{https://data.nrao.edu/portal/}). Relevant data files, analysis scripts and custom codes are archived at Zenodo (\dataset[DOI: 10.5281/zenodo.13939362]{https://doi.org/10.5281/zenodo.13939362}).
\end{acknowledgments}

\begin{contribution}
M.C.B. and F.C.Z. contributed equally to this work. M.C.B. and F.C.Z. designed the observing campaign and led the paper writing, with input from all authors. M.C.B. was the Principal Investigator of the \ixpe, VLT, and VLA proposals. F.C.Z. obtained \nicer\ observations. S.Ca. obtained \swift\ observations. M.C.B. performed the reduction and analysis of the VLT data and wrote the corresponding text. F.C.Z. performed the reduction and analysis of the \nicer\ and \swift\ data, developed the procedure to select emission modes in the \ixpe\ data, and wrote the corresponding text. A.D.M. and F.L.M. performed the reduction and analysis of the \ixpe\ data and wrote the corresponding text. A.P. carried out the timing analysis of the data and wrote the corresponding text. A.K.H. performed  the reduction and analysis of the VLA data and wrote the corresponding text. S.G. and S.E.M. assisted with preparing the VLA proposal. A.P. provided SED data points from archival observations. All authors contributed to the discussion of the presented results and provided comments on the manuscript.
\end{contribution}

%

\facilities{\ixpe, \nicer, \swift\ (XRT and UVOT), VLA, VLT:Antu, ADS, HEASARC}


\software{
ASTROPY v.7.0.1, a community-developed core Python package for Astronomy \citep{astropy:2013,astropy:2018,astropy:2022}; 
CASA v.6.5 \citep{casa}; 
DAOPHOT \citep{Stetson1987};
DUST$_-$EXTINCTION v.1.6 \citep{Gordon2024};
HEASOFT v.6.34 \citep{heasoft14}; 
IPYTHON v.9.0.2 \citep{Perez2007};
IRAF v.2.18, distributed by the National Optical Astronomy Observatory, which is operated by the Association of Universities for Research in Astronomy, Inc., under a cooperative agreement with the National Science Foundation (\url{https://iraf-community.github.io/});
IXPEOBSSIM v.31.0.3 \citep{Baldini2022}, documented at \url{https://ixpeobssim.readthedocs.io/en/latest/};
MATPLOTLIB v.3.10.1, a Python library for publication-quality graphics \citep{hunter07}; 
NICERDAS v.13;
NUMPY v.2.2.4 \citep{harris20}; 
PANDAS v.2.2.3 \citep{mckinney2010};
PINT v.1.1.1 \citep{Luo2021};
QUARTICAL v.0.2.3 \citep{quartical};
SAOImageDS9 v.8.6, a tool for data visualization supported by the Chandra X-ray Science Center (CXC) and the High Energy Astrophysics Science Archive Center (HEASARC) with support from the James Webb Space Telescope Mission office at the Space Telescope Science Institute for 3D visualization \citep{joye03}; 
SCIPY v.1.15.2 \citep{scipy20}; 
Swift Reduction Package (SRPAstro) v.4.9.0 (\url{https://pypi.org/project/SRPAstro/});
WSClean v3.4 \citep{Hogbom1974};
XRONOS v.5.22 \citep{stella92}; 
XSPEC v.12.14.1 \citep{arnaud96}.
}



\appendix

\section{Selection of Emission Modes in the \ixpe\ Dataset}
\label{sec:ixpe_modes}
The top panel of Fig.\,\ref{fig:timeseries_X} shows the time series for the \ixpe\ observations, whereas the middle and bottom panels show the six time intervals where there was overlap between \ixpe\ and \nicer\ observations, totaling $\simeq$2\,ks. In each interval, the average background intensity in the \nicer\ dataset was $\simeq$10--15\% of the total emission (source plus background), except for the last time segment, where the emission was consistently faint, and the background intensity accounted for $\simeq$30\% of the total emission. Nonetheless, during these overlapping time intervals, we did not observe any significant variation in the background level possibly correlated with mode-switching episodes, except for the fifth interval, where a slightly enhanced background level was detected at the exit of a low mode. This increase in the count rate should be approached with caution. Overall, we detected four low modes in the \nicer\ data that were covered by \ixpe: one in the second time segment, one in the third, one in the fifth, and one throughout the entire duration of the sixth segment.

\begin{figure*}
\begin{center}
\includegraphics[width=0.75\textwidth]{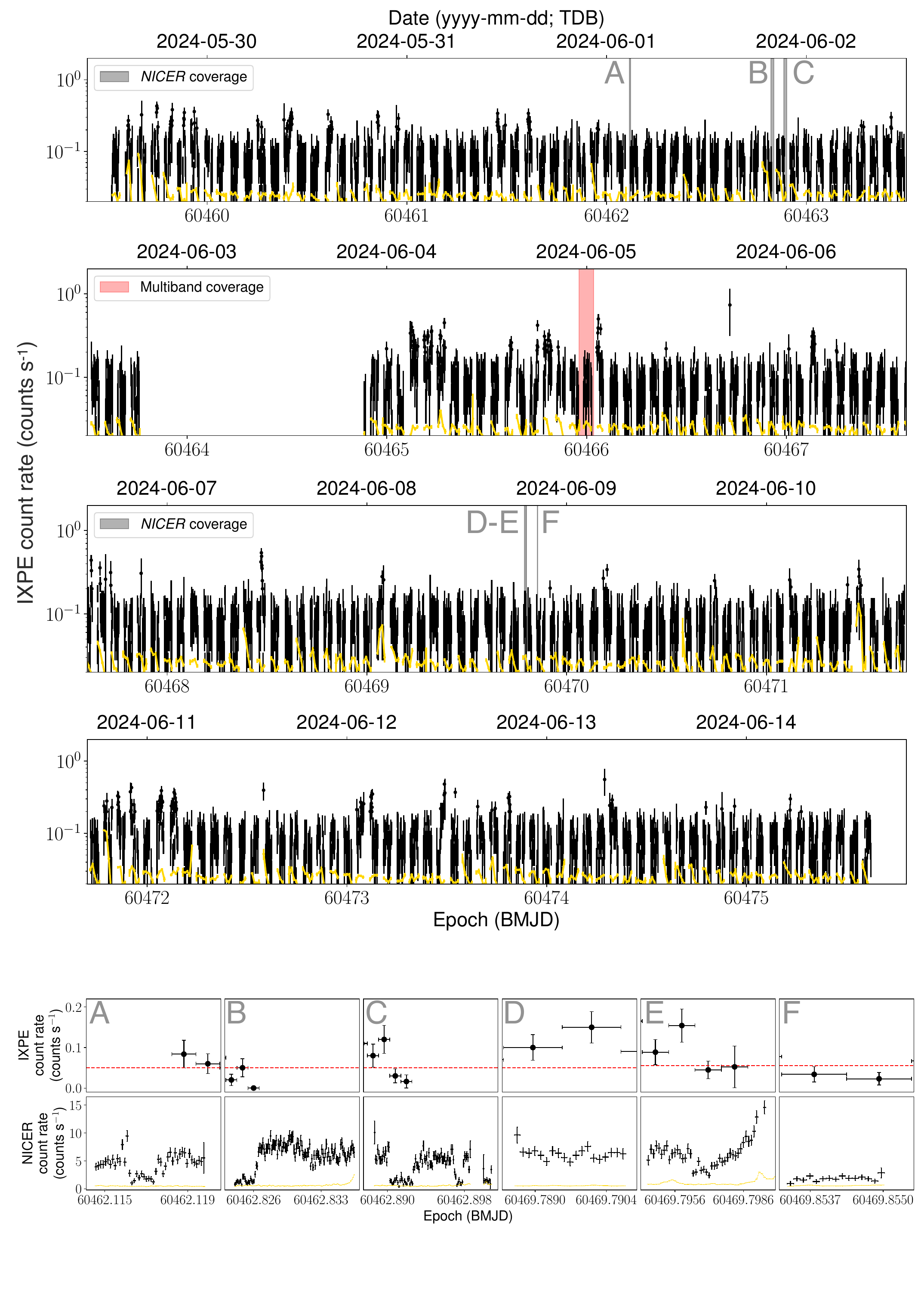}
\vspace{-1cm}
\caption{Top: \ixpe\ time series extracted from the source region and binned at a 100-s resolution (black) and from the background region and binned at 1\,ks (orange). Both time series were extracted by combining data from the three detector units, with the background time series rescaled to account for the area differences between the source and background extraction regions. The gaps in the time series correspond to time intervals when J1023 was obscured by the Earth. The vertical gray stripes mark the six-time intervals of the simultaneous \nicer\ observations, while the blue-shaded area represents the time interval covered by VLT and VLA observations in the optical and radio bands. Bottom: \ixpe\ and \nicer\ time series for each overlapping time interval, with the \nicer\ data binned at 10\,s. The horizontal red dashed lines in the middle panels indicate the count rate threshold adopted to differentiate between high and low mode episodes in the \ixpe\ dataset, while the cyan dashed lines in the bottom panels depict the \nicer\ background level estimated using the \texttt{SCORPEON} model. The final part of the fifth time segment of the \nicer\ dataset is characterized by enhanced background level. Therefore, caution should be exercised in interpreting the corresponding increase in the count rate. BMJD stands for Barycentric Modified Julian Date, TDB for Barycentric Dynamical Time.}
\label{fig:timeseries_X}
\end{center}
\end{figure*}

To select the appropriate thresholds for the count rates in order to select the distinct emission modes in the \ixpe\ data, we proceeded as follows. Over the past decade, the net count rate observed in all previous observations of J1023 carried out by the European Photon Imaging Cameras (EPIC) on board \xmm\ decreased by a factor of approximately six during the low mode compared to the average net count rate (e.g., \citealt{Bogdanov15,cotizelati18a}). The average count rate of J1023 measured by \ixpe\ after combining data from the three DUs is $\simeq$0.09\,counts\,s$^{-1}$, with about 30\% of this rate contributed by the background level. Consequently, during low-mode episodes, typically only 0 or 1 net photon from the source, plus 3--4 photons from the background, would be detected by \ixpe\ in 100-s time bins. In fact, during the low modes covered by \nicer, the highest count rate measured in the \ixpe\ time series binned at 100\,s is 0.05\,counts\,s$^{-1}$ (summing up the source and background contributions). We then classified time intervals in the \ixpe\ light curves as low modes whenever the count rate in the 100-s binned time series was less than or equal to 0.05\,counts\,s$^{-1}$. This criterion resulted in the selection of about 182\,ks of low-mode data, which corresponds to about 27\% of the total exposure time. This value is consistent with the fraction of time that J1023 was detected in the low mode in previous observations \citep{Archibald15,Bogdanov15,cotizelati18a}.

To differentiate between high and flaring modes, we extracted the distribution of count rates and modeled it using Gaussian mixture models. We then classified a 100-s time bin as a flare mode if its count rate was at least 4 standard deviations above the mean count rate of the Gaussian distribution representing the high mode, which corresponds to a threshold of $\approx$0.2\,counts\,s$^{-1}$ (Fig.\,\ref{fig:ixpe_cr_distrib}). This approach resulted in the selection of $\approx$20\,ks of flaring-mode data, which represents $\simeq$3\% of the total duration of the \ixpe\ observations. This is again consistent with the fraction observed in previous observations \citep{Archibald15,Bogdanov15,cotizelati18a}.

\begin{figure}
\begin{center}
\includegraphics[width=0.48\textwidth]{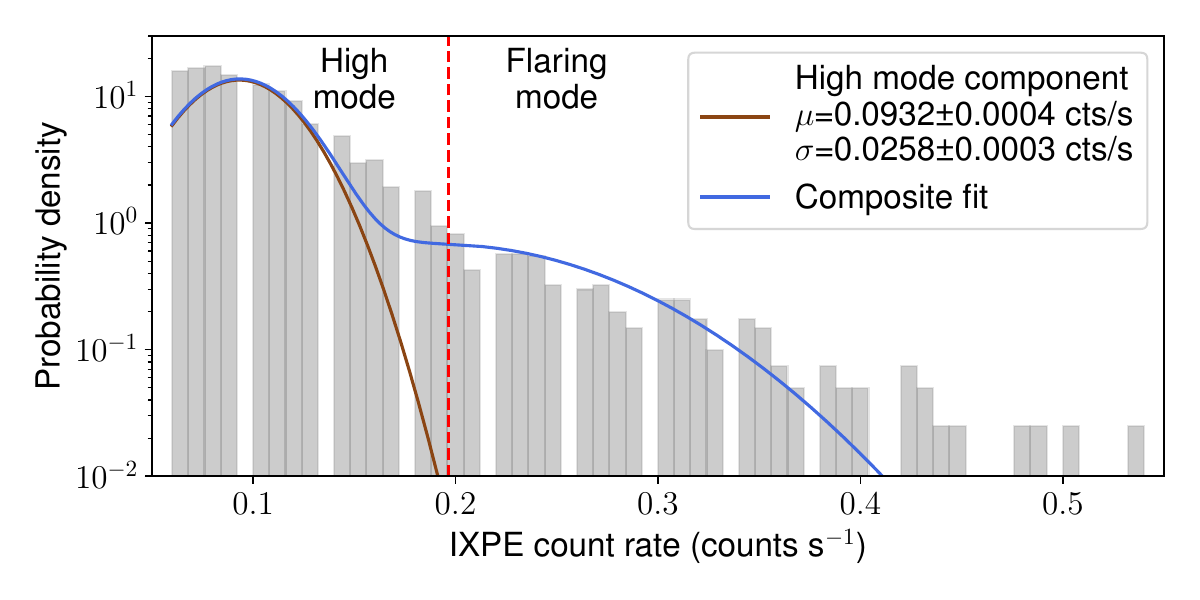}
\caption{Probability density function of \ixpe\ count rates and its modeling. The gray filled bins represent the distribution of count rates extracted from time series binned at 100\,s. The gaps between some of the bins in the distribution result from the limited number of photon counts detected by \ixpe\ in each time bin. The distribution was fit with a combination of three Gaussian functions. The blue line represents the probability density function (PDF) of the composite model that best fits the observed data. The brown curve represents the PDF of the Gaussian component associated with the high mode. The mean ($\mu$) and standard deviation ($\sigma$), along with their 1-$\sigma$ uncertainties, are provided in the legend. The vertical red dashed line marks the threshold adopted to identify the flaring mode, set at 0.2\,counts\,s$^{-1}$, which corresponds to 4$\sigma$ above the mean of the high mode component. The vertical axis is plotted on a logarithmic scale to highlight the tail of the distribution at higher count rates in the flaring mode.}
\label{fig:ixpe_cr_distrib}
\end{center}
\end{figure}

\section{Timing Analysis of the \ixpe\ Dataset}
\label{sec:timing_app}
The top panel of Fig.\,\ref{fig:pol_phases} shows the \ixpe\ background-subtracted pulse profile. We calculate the fractional RMS variability amplitude ($F_{\rm var}$) of the pulsed signal following the method by \cite{vaughan03} and obtain $F_{\rm var}$ = 8.1\%$\pm$0.8\%. This value is consistent within the uncertainties with those derived in previous studies \citep{Archibald15,jaodand16,papitto19}.

The spin frequency measured from \ixpe\ data is compatible within the uncertainties with the value predicted from the quasi-coherent timing solution derived from optical data by \cite{Burtovoi2020} ($\nu_{\rm exp, B20}=592.421467042(7)$\,Hz), who found a spin down rate very close to the value observed when J1023 was detected as a radio pulsar before the state transition occurred in 2013. On the other hand, it differs by 160$\pm$90\,nHz from the value predicted by \cite{jaodand16} ($\nu_{\rm exp,J16}=592.421466890(3)$\,Hz), who reported a spin-down rate about 20\% faster (Fig.\,\ref{fig:efsearch}). Even though this difference is still compatible within a 3$\sigma$ c.l., our measurement tends to favor a scenario in which the rotational evolution of the pulsar in the subluminous X-ray state has remained largely unchanged compared to its behavior during the radio pulsar state.

\begin{figure}[!h]
\begin{center}
\includegraphics[width=0.48\textwidth]{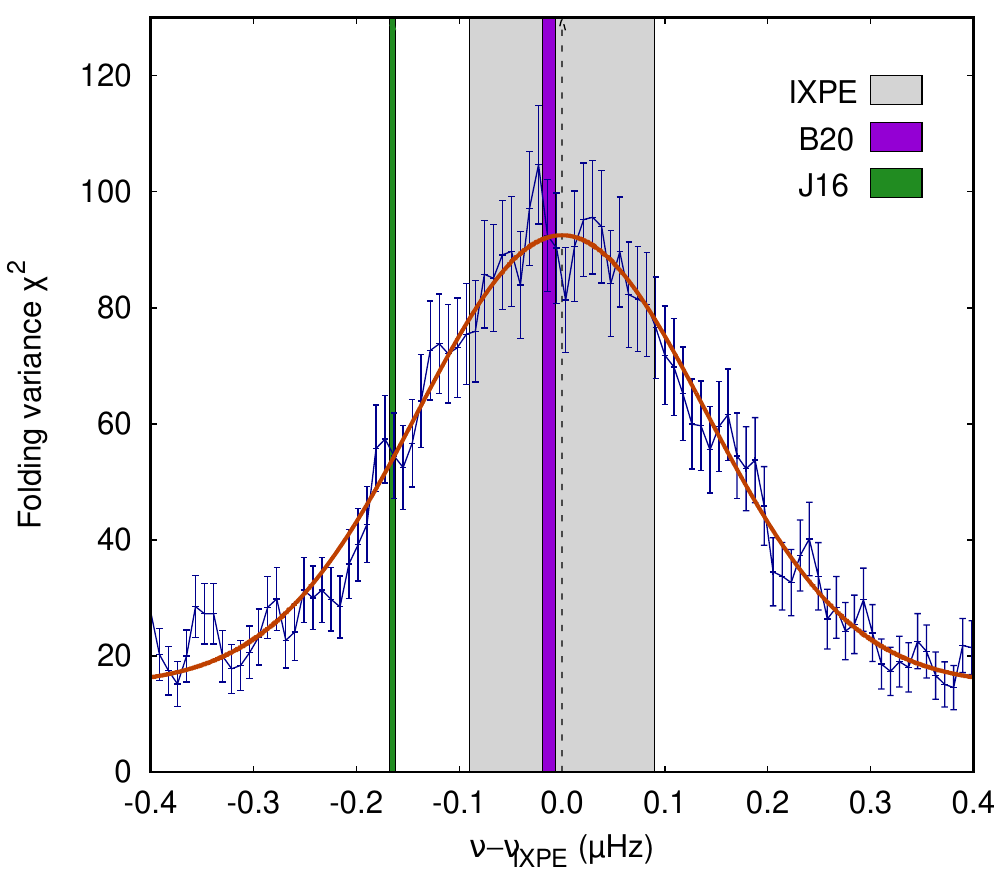}
\caption{Variance distribution for the coherent pulsations of J1023 detected by \ixpe. The distribution is derived from an epoch folding search of \ixpe\ data filtered for the high mode in the 2--6\,keV energy range. The folding variance is plotted against the frequency deviation from the \ixpe-measured spin frequency, $\nu_{\rm IXPE}$ = 592.42146705(9)\,Hz, marked with a vertical dashed line. The red solid curve indicates the best-fitting Gaussian function to the pulse variance distribution. The shaded areas indicate the uncertainty regions around the spin frequency measured by \ixpe\ (gray) or those predicted by extrapolating the solutions reported by \cite{Burtovoi2020} (B20; purple) and \cite{jaodand16} (J16; green) to the epoch of the \ixpe\ observation.}
\label{fig:efsearch}
\end{center}
\end{figure}

\section{X-ray Polarimetric Properties}
\label{sec:pol_app}
\subsection{Model-independent Analysis}
\label{sec:pol_mi_app}
We first perform a model-independent polarimetric analysis of the \ixpe\ data using the formalism outlined by \cite{Kislat2015} implemented in the \texttt{ixpeobssim} package \citep{Baldini2022} under the \texttt{pcube} algorithm in the \texttt{xpbin} routine using the unweighted approach \citep{DiMarco2022}. We compute the normalized, background-subtracted Stokes parameters $q_X = Q_X/I_X$ and $u_X = U_X/I_X$, where $Q_X$, $U_X$ and $I_X$ are the Stokes parameters. From these, we derive the polarization degree $P_X = \sqrt{q_X^2 + u_X^2}$ and polarization angle (also referred to as the electric vector position angle) $PA_X = \frac{1}{2}\arctan\left(u_X/q_X\right)$, measured counterclockwise from the North celestial pole toward the East. Hereafter, uncertainties are given at the 68.27\% c.l. unless noted otherwise.

In the 2--6\,keV energy band, we obtain an MDP of 11\% at a 99\% c.l. and $P_X = (7\pm4)$\%. 
The polarization in the 2--3\,keV energy range is consistent with zero within $1\sigma$, while low-significance polarization is consistently detected across the 3--4, 4--5, and 5--6\,keV energy ranges (Fig.\,\ref{fig:Stokes_X}). In the 3--6\,keV range, we find an MDP of 15\% and measure a polarization degree that exceeds this threshold, $P_X = (16\pm5)$\%, with $PA_X = -7\deg\pm9\deg$. To assess the significance of this detection, we use the ratio of the polarization degree to its standard deviation, $P_X / \sigma_{P_X}=3.2$, as the primary indicator of statistical significance\footnote{\url{https://heasarc.gsfc.nasa.gov/docs/ixpe/analysis/IXPE_Stats-Advice.pdf}}. Using the $\chi^2$ distribution for two dofs, we calculate a confidence level (c.l.) of 99.4\%. Considering the 2--3\,keV and 3--6\,keV energy bins as independent data sets, we tested the combined significance of the polarization detection against the null hypothesis (i.e., zero polarization in every bin). This analysis yielded a polarization detection at a 96.1\% c.l., corresponding to a $3.87\times10^{-2}$ probability for the null hypothesis.

\begin{figure}
\begin{center}
\includegraphics[width=0.45\textwidth]{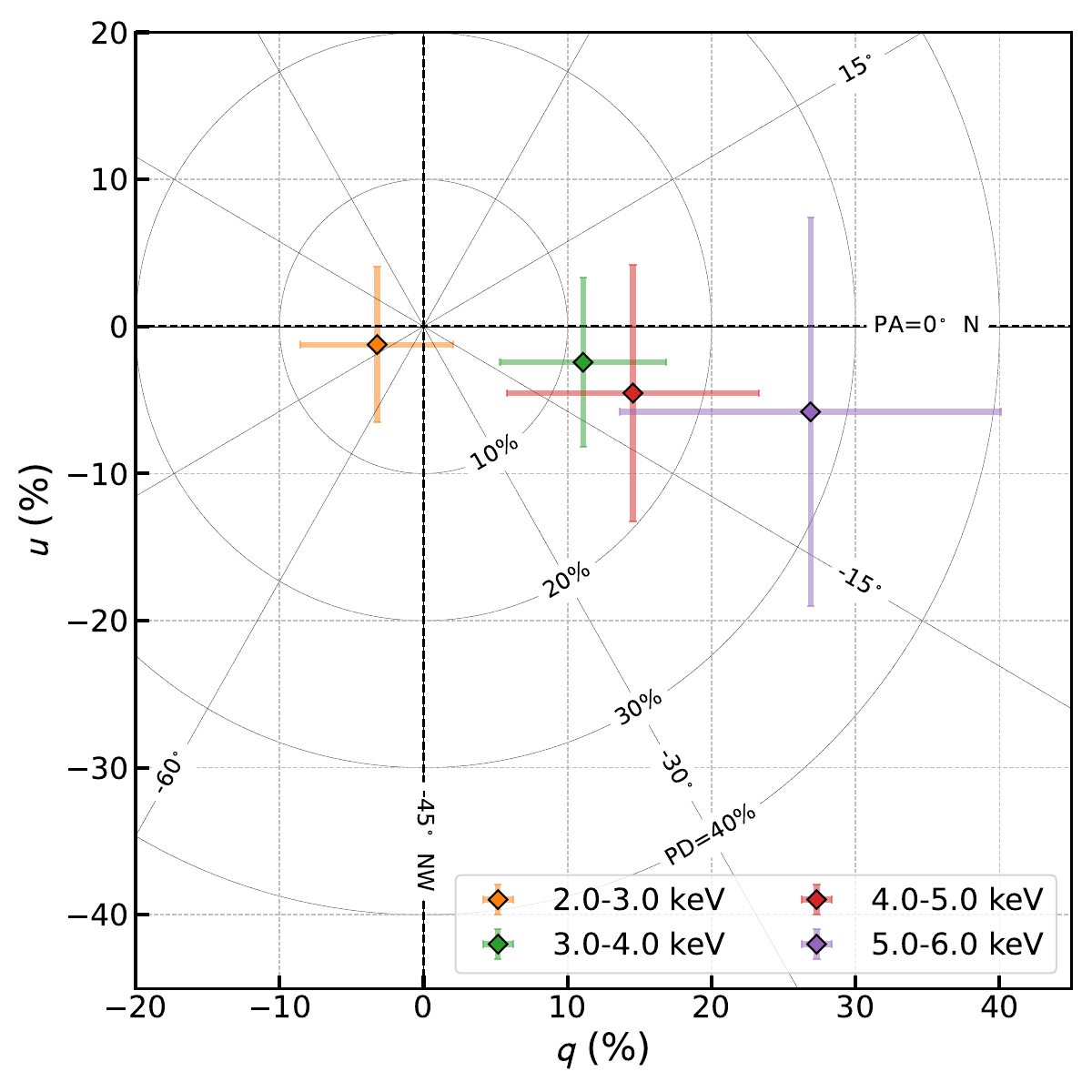}
\vspace{-0.2cm}
\caption{Normalized, background-subtracted Stokes parameters $q$ and $u$ for the average X-ray emission. The results are derived from the analysis of the combined data from all three DUs with no selection of the modes. The Stokes parameters were grouped into four bins with sizes of 1\,keV across the 2--6\,keV energy band. Each diamond represents the mean value, with error bars indicating the 1$\sigma$ standard deviation. The circles give the contours of constant polarization degree while the radial lines correspond to constant polarization angle.}
\label{fig:Stokes_X}
\end{center}
\end{figure}

Applying the same analysis to high-mode data, we obtain an MDP of 12\% in the 2–6\,keV range, with $P_{\rm X,H} = (7\pm4)$\% and $PA_{\rm X,H} = -9\deg\pm18\deg$. The energy-resolved analysis corroborates the results from the entire data set. 
In the 3--6\,keV range, we find an MDP of 16\% and $P_{\rm X,H} = (13\pm5)$\%, with $PA_{\rm X,H} = -1\deg\pm11\deg$. 

For low-mode and flaring-mode data, the significance was too low for a reliable polarimetric analysis. For the low mode, we obtained an MDP of 50\% and $P_{\rm X,L} < 26$\% at a 90\% c.l.. For the flaring mode, we derived an MDP of 30\% and $P_{\rm X,F} < 28$\% at a 90\% c.l..

\subsection{Spectro-polarimetric Analysis}
\label{sec:pol_md_app}
To enhance the significance of the polarimetric measurements, we conducted a spectro-polarimetric analysis of the data as outlined by \cite{Strohmayer2017} and adopted the weighted analysis method prescribed by \cite{DiMarco2022}, where each photoelectron track recorded by the DUs is assigned a weight based on its ellipticity. We extracted the $Q_X$, $U_X$ and $I_X$ spectra for each DU using the \texttt{xpbin} tool's \texttt{PHA1} algorithms in \texttt{ixpeobbsim}, applying a constant energy binning of 200\,eV. We assigned the response matrices \texttt{20240701}$_-$\texttt{alpha075} available in the calibration database, which are appropriate for the time span covered by the observations, and computed modulation response functions and ancillary response files using \texttt{ixpecalcarf}. 

Previous observations have shown that the X-ray spectrum of J1023 during its high mode is well described by an absorbed power-law model (e.g., \citealt{Bogdanov15,cotizelati18a,campana19}). Hence, the adopted model includes a power-law component corrected for interstellar medium effects using the T\"ubingen-Boulder model \citep{wilms00} with photoelectric cross-sections by \cite{verner96}, convolved with an energy-independent polarization model.
To account for potential discrepancies between the effective areas of the different DUs, we also included a cross-normalization constant, which was fixed at unity for DU1 to serve as the reference, and allowed to vary for DU2 and DU3.
Thus, the final model employed within the \texttt{xspec} spectral fitting package \citep{arnaud96} is \texttt{const $\times$ TBabs $\times$ powerlaw $\times$ polconst}. We fit this model simultaneously to the Stokes energy spectra of all DUs across the 2--6, 2--3 and 3--6\,keV energy ranges using the standard forward-folding, maximum likelihood procedure implemented in \texttt{xspec}. 

Observations over the past decade have shown that the X-ray spectral shape of J1023 has remained remarkably stable throughout this time span \citep{Bogdanov15,cotizelati18a,campana19}. Hence, during the fitting process, we fixed the absorption column density ($N_{\rm H}$) and power-law photon index ($\Gamma_X$) at the values measured in previous works: $N_{\rm H}=2.8\times10^{20}$\,cm$^{-2}$, $\Gamma_X=1.69$. All other parameters were allowed to vary.
The results are summarized in Table\,\ref{tab:X_pol} and are consistent with those derived from the model-independent analysis. Figure\,\ref{fig:polar_plot} shows the confidence contours for the measurements of $P_{\rm X,H}$ and $PA_{\rm X,H}$ obtained over distinct energy bands, drawn using the \texttt{steppar} command in \texttt{xspec}. 

It is noteworthy that previous analyses of the distributions of mode durations extracted from observations carried out by \xmm\ EPIC \citep{Archibald15} show that $\simeq$40\% of low-mode episodes last less than 100\,s. This implies that our method for selecting the emission modes inherently misses short low-mode episodes totaling $\simeq$10\% of the total duration of the \ixpe\ observations. According to \cite{Baglio2023}, the emission during the low mode is expected to be polarized to a lower extent compared to the high mode due to the much smaller contribution of the emission from the boundary region. Even in the most extreme case where the low-mode emission is fully unpolarized, the X-ray polarization degree during the high mode selected using our criteria would be enhanced by a factor of $1+\frac{0.1\times CR_{\rm low}}{CR_{\rm high}} \simeq 1.01$ \citep{DiMarco2023}. Here, $CR_{\rm low}$ and $CR_{\rm high}$ represent the background-subtracted count rates in the \ixpe\ band in the low and high modes, respectively. This small increase, equivalent to about 0.2\% in the polarization degree, is within the margin of the measurements uncertainties, meaning that as a matter of fact the intrinsic X-ray polarization degree of the high-mode emission remains unchanged within the limits of our observations.

\begin{figure}[h]
\begin{center}
\includegraphics[width=0.45\textwidth]{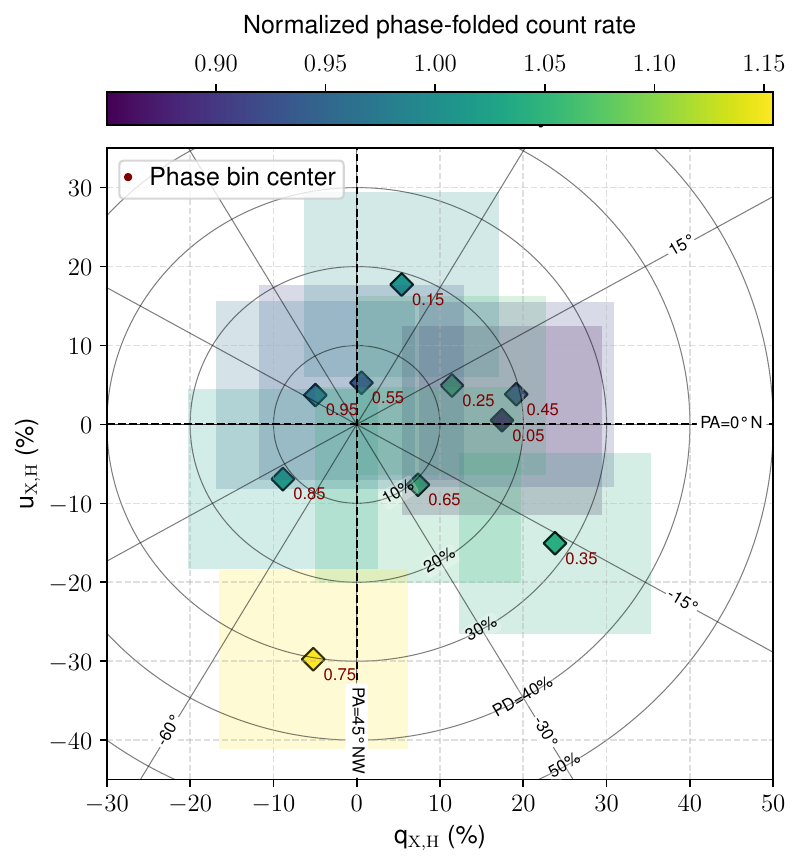}
\hspace{1cm}
\vspace{-0.3cm}
\caption{Normalized, background-subtracted Stokes parameters in the high mode in the 2-6\,keV energy in distinct pulsar rotational phase bins. The results are derived from an unbinned analysis of the combined data from all three DUs. Each diamond represents the mean value for a phase bin, with the center of each phase bin highlighted in bordeaux. The corresponding shaded circular area shows the $1\sigma$ standard deviation. The circles give the contours of constant polarization degree while the radial lines correspond to constant polarization angle. Both diamonds and square areas are color-coded based on the normalized background-subtracted count rate in the corresponding pulse profile phase bin.} 
\label{fig:q_u_planes}
\end{center}
\end{figure}

\begin{figure*}
\begin{center}
\includegraphics[width=0.38\textwidth]{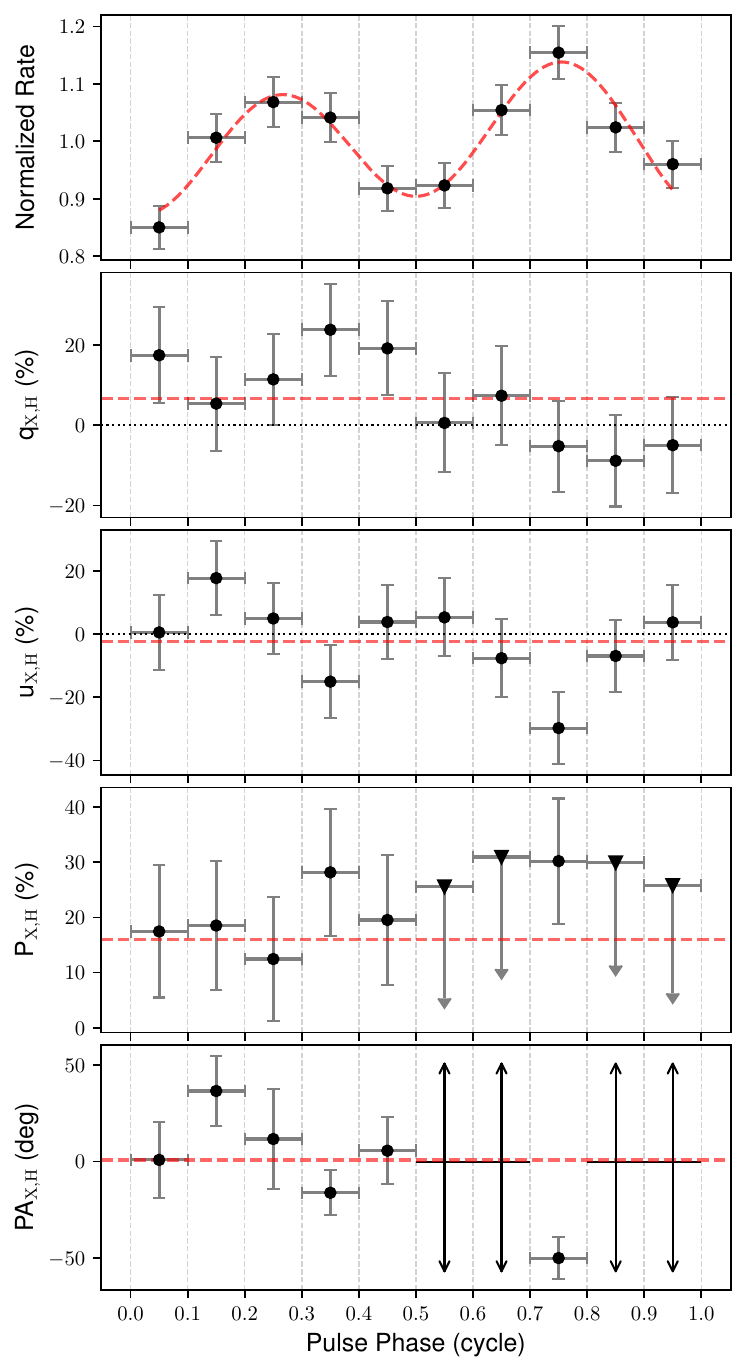}
\hspace{1cm}
\includegraphics[width=0.38\textwidth]{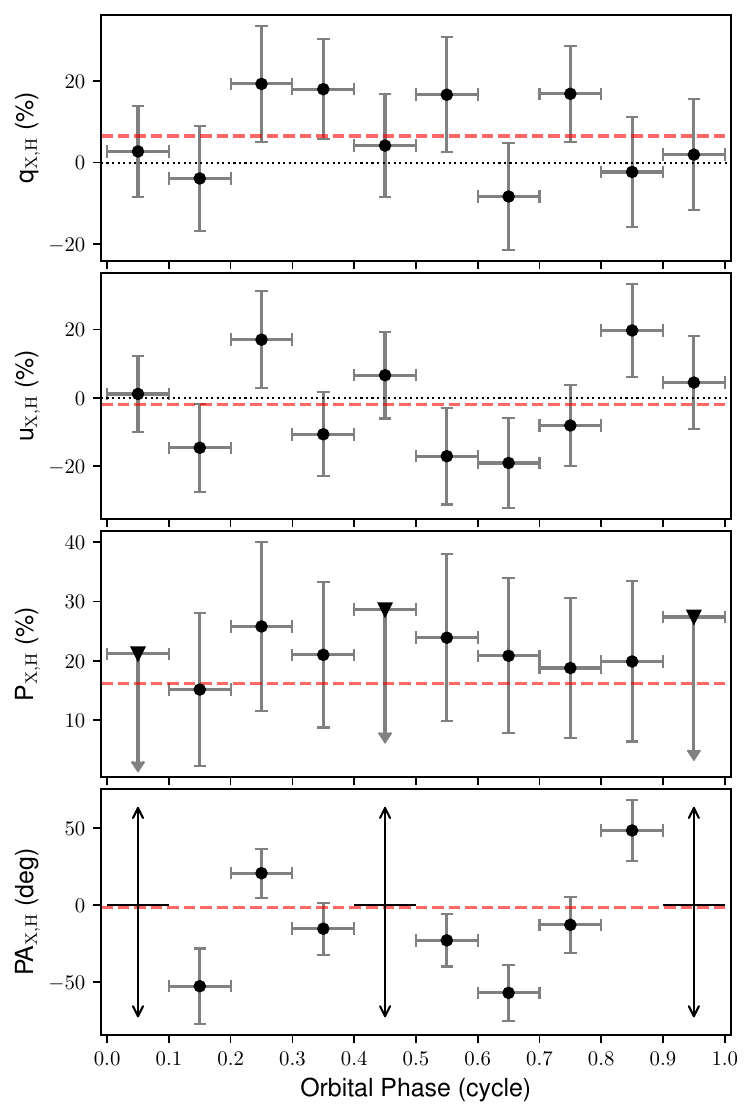}
\vspace{-0.3cm}
\caption{Rotational (left) and orbital (right) phase dependence of the 2--6\,keV normalized flux and polarization properties of J1023 in the high mode. 
From top to bottom, the left panel shows the  pulse profile normalized to the average background-subtracted count rate as well as the normalized Stokes parameters $q_X$ and $u_X$, the $P_{\rm X,H}$, and the $PA_{\rm X,H}$ sampled in 10 phase bins as a function of the rotational phase and after background subtraction. From top to bottom, the right panel shows the normalized Stokes parameters $q_X$ and $u_X$, the $P_{\rm X,H}$, and the $PA_{\rm X,H}$ sampled in 10 phase bins as a function of the orbital phase and after background subtraction (here phase 0 corresponds to the time of passage of the pulsar through the ascending node of the orbit). In both panels, upper limits are indicated with arrows and are reported at the 90\% c.l. for bins compatible with null values at the 68\% c.l.. The $PA_{\rm X,H}$ values are unconstrained for bins where upper limits on the $P_{\rm X,H}$ have been derived. The red dashed line in the top left panel indicates the best-fitting model comprising the fundamental and first harmonic components of the spin signal, whereas the red horizontal dashed lines in all other panels indicate the mean of the plotted polarization parameters. $q_X$ and $u_X$ are statistically consistent with remaining constant across the phases (see text for more details).} 
\label{fig:pol_phases}
\label{fig:pol_phases_orb}
\end{center}
\end{figure*}

\subsection{Phase-Resolved Analyses}
\label{sec:ppps_app}
We conducted both rotational and orbital phase-resolved analyses of the X-ray polarimetric properties. Using the \texttt{photonphase} tool in the \texttt{PINT} software package \citep{Luo2021}, we computed the rotational and orbital phases of each photon collected in the high mode in the 2--6\,keV energy range by \ixpe, based on the timing model derived from the \ixpe\ data analysis. For each analysis, we split the data into 10 phase bins, and calculated the Stokes parameters for each bin using the \texttt{pcube} algorithm within \texttt{ixpeobssim}, following the same model-independent procedure described above and applying background subtraction.

Figure\,\ref{fig:q_u_planes} shows the Stokes parameters in distinct pulsar rotational phases.
The left panel of Figure\,\ref{fig:pol_phases} displays the Stokes parameters, $P_{\rm X,H}$ and PA$_{\rm X,H}$ as a function of the rotational phase. 
We do not measure statistically significant (anti-)correlations between the normalized rate and the Stokes parameters. Specifically, Spearman rank and Kendall tau correlation analyses between the normalized rate and $q_{X,H}$ yielded coefficients of -0.25 (p-value = 0.49) and -0.16 (p-value = 0.60), respectively. Similarly, correlation analyses between the normalized rate and $u_{X,H}$ resulted in coefficients of -0.47 (p-value = 0.17) and -0.33 (p-value = 0.22), respectively. Constant fits to the $q_{X}$ and $u_{X}$ values produced $\chi^2$ values of 8.5 and 11.8 for 9 dof, respectively. The right panel of Figure\,\ref{fig:pol_phases_orb} shows the Stokes parameters, $P_{\rm X,H}$ and PA$_{\rm X,H}$ as a function of the orbital phase. Similarly, no significant variability was observed in $q_{X}$ and $u_{X}$, with constant fits yielding $\chi^2$ values of 5.6 and 9.8 for 9 dof, respectively.

\section{Optical Properties}
\label{sec:opt_app}
All VLT observations resulted in significant detections (above 3$\sigma$), with a mean linear polarization level of $P_{\rm opt}=(1.38\pm 0.04)\%$ (1$\sigma$). This value aligns with previous results obtained in the same band \citep{baglio16,Baglio2023}, although it is slightly higher. The polarization angle is well-constrained, with a mean value of  $PA_{\rm opt}=-4.1\deg\pm0.7\deg$  (1$\sigma$; measured after correcting for the polarization angle of the standard star). This value is consistent with that reported by \cite{Baglio2023}, although our measurements are significantly more precise. 

In addition to the polarimetric analysis, it is also possible to extract and calibrate fluxes to build a light curve. If we assume the flux loss is negligible, the sum of the intensities measured in both the ordinary and extraordinary beams for each image gives the total flux from the target. Therefore, we summed the ordinary and extraordinary fluxes for each image of J1023, as well as for two isolated field stars with R2 magnitudes listed in the USNO B1 catalogue \citep{monet03}. The resulting flux was then calibrated against the flux of the reference stars.

Figure\,\ref{fig:lcurves} shows the time evolution of the magnitude, P$_{\rm opt}$ and PA$_{\rm opt}$. To quantify the intrinsic variability amplitude of the light curve, we computed the excess variance $\textit{ExcessVar} = \sigma_{\text{m}}^2 - \overline{\sigma_{\text{m,err}}^2}$, where $\sigma_{\text{m}}$ is the standard deviation of the magnitudes and $\overline{\sigma_{\text{m,err}}^2}$ is the mean square measurement error in magnitudes. We obtain \textit{ExcessVar} $\simeq$ 0.02\,mag$^2$. Converting this magnitude variability to flux variations, we derive an intrinsic fractional variability of $\simeq$13\%, indicating moderate variability.

During the low mode episode, we observe an increase in the optical emission by a factor of 1.3, which contrasts with the behavior seen in the X-ray and UV bands. The origin of this optical increase is unclear, but it may be related to the broadening or enhanced prominence of hydrogen H$\alpha$ and/or three helium He I emission lines. These lines, possibly originating from regions within the accretion disk or from mass outflows, fall within the wavelength range of the $R$ filter of FORS2 (572.5--737.5\,nm). A recent high-time-resolution optical spectroscopic study of J1023 using the OSIRIS spectrograph mounted on the \emph{Gran Telescopio Canarias} reported significant variability in the equivalent width and full width at half maximum of these lines, such as changes by a factor of two over minute timescales for the H$\alpha$ line, with no clear correlation with the flux of the optical continuum \citep{Messa2024} (see also \citealt{shahbaz19, hakala18} for further evidence of highly variable emission line profiles). In contrast, no prominent emission lines are expected within the wavelength range of the UVM2 filter of the \swift\ UVOT, as shown in the UV spectrum by \cite{Hernandez2016}. We emphasize that if the optical emission lines originate from an outflow, this observation would lend support to scenarios suggesting stronger outflows during the low mode \citep{Veledina19, Baglio2023}, which also result in increased radio emission. However, since similar variability in the optical light curve is also observed in other instances during the X-ray high mode, it is not possible to draw a definitive conclusion.

P$_{\rm opt}$ exhibits moderate time variability, with a Median Absolute Deviation (MAD) of 0.23\% and a normalized Excess Variance (NEV) of 0.05, indicating some intrinsic fluctuations. Its skewness is 0.58, suggesting a slight right-skewed distribution and indicating that deviations above the mean are more frequent. On the other hand, PA$_{\rm opt}$ shows greater variability, with a MAD of 3.6\deg, a significantly higher NEV of 12.7, and a skewness of -0.23, pointing to a slight left-skewed distribution where deviations below the mean are more common. 
The opposite trends in the skewness of P$_{\rm opt}$ and PA$_{\rm opt}$ are primarily due to their anticorrelated variability during the second peak of optical intensity (Figures\,\ref{fig:lcurves} and \ref{fig:Stokes_opt_ev}). We measure $P_{\rm opt,H}=(1.41\pm 0.04)\%$ and $PA_{\rm opt,H}=-3.9\deg\pm0.7\deg$ for the high mode; $P_{\rm opt,L}=(0.97\pm 0.13)\%$ and $PA_{\rm opt,L}=-7.6\deg\pm4.0\deg$ for the low mode. Consistent with previous findings by \cite{Baglio2023}, these measurements suggest a slight decrease in polarization degree when switching from the high to the low mode, while the polarization angle shows no significant variation within the uncertainties.

Figure\,\ref{fig:Stokes_opt_ev} shows the time series of the Q$_{\rm opt}$ and U$_{\rm opt}$ Stokes parameters. The two parameters exhibit distinct variability patterns. Q$_{\rm opt}$ exhibits relatively low variability, with a MAD of 0.0013 and a NEV of 0.03, indicating minimal intrinsic fluctuation. Its skewness is -0.33, suggesting a slightly left-skewed distribution with more frequent deviations below the mean. U$_{\rm opt}$ shows more pronounced variability, with a MAD of 0.0016 and a significantly higher NEV of 1.8, indicating substantial intrinsic fluctuations. The skewness is 0.04, indicating an almost symmetric distribution, with a very slight tendency for deviations above the mean. These results suggest that U$_{\rm opt}$ is more variable and less symmetric than Q$_{\rm opt}$. 
 
Figure\,\ref{fig:Stokes_opt_plane} also shows the $Q_{\rm opt}$ and $U_{\rm opt}$ Stokes parameters of J1023 on the $Q_{\rm opt}-U_{\rm opt}$ plane for the whole dataset, including high modes (filled dots) and the low mode (filled diamond).
The data points are color-coded according to the time they were recorded. Despite some intrinsic variation, the $Q_{\rm opt}$ and $U_{\rm opt}$ values cluster around a common point, and there is no clear distinction between the high and low modes.

\begin{table}
\caption{Optical polarimetric measurements. All uncertainties are at 1$\sigma$ c.l.. BMJD stands for Barycentric Modified Julian Date.}
\label{tab:opt_pol}
\centering
\resizebox{\columnwidth}{!}{
\begin{tabular}{ccccc}
\hline\hline
Epoch       & Q$_{\rm opt}$  & U$_{\rm opt}$  & P$_{\rm opt}$  & PA$_{\rm opt}$ \\
(BMJD)      &         &         & (\%)    & ($^\circ$) \\
\hline
60465.9887  & 0.0131 $\pm$ 0.0014 & -0.0009 $\pm$ 0.0014 & 1.43 $\pm$ 0.13 & 2.0 $\pm$ 2.8  \\
60465.9915  & 0.0127 $\pm$ 0.0014 &  0.0009 $\pm$ 0.0014 & 1.44 $\pm$ 0.13 & -0.1 $\pm$ 2.7  \\
60465.9943  & 0.0101 $\pm$ 0.0014 &  0.0005 $\pm$ 0.0014 & 1.01 $\pm$ 0.13 & 2.8 $\pm$ 3.8  \\
60465.9971  & 0.0074 $\pm$ 0.0014 & -0.0037 $\pm$ 0.0014 & 0.97 $\pm$ 0.13 & -7.6 $\pm$ 4.0  \\
60465.9999  & 0.0129 $\pm$ 0.0014 & -0.0012 $\pm$ 0.0014 & 1.51 $\pm$ 0.14 & -2.4 $\pm$ 2.6 \\
60466.0027  & 0.0090 $\pm$ 0.0014 & -0.0044 $\pm$ 0.0012 & 1.03 $\pm$ 0.13 & -4.7 $\pm$ 3.2 \\
60466.0055  & 0.0131 $\pm$ 0.0014 & -0.0069 $\pm$ 0.0013 & 2.12 $\pm$ 0.13 & -12.3 $\pm$ 1.9 \\
60466.0083  & 0.0113 $\pm$ 0.0014 & -0.0041 $\pm$ 0.0014 & 1.60 $\pm$ 0.13 & -11.4 $\pm$ 2.1 \\
60466.0112  & 0.0101 $\pm$ 0.0014 & -0.0018 $\pm$ 0.0014 & 1.12 $\pm$ 0.14 & -0.3 $\pm$ 3.5  \\
60466.0140  & 0.0129 $\pm$ 0.0014 &  0.0021 $\pm$ 0.0014 & 1.57 $\pm$ 0.13 & 5.9 $\pm$ 2.5  \\
60466.0168  & 0.0138 $\pm$ 0.0014 & -0.0028 $\pm$ 0.0014 & 1.64 $\pm$ 0.14 & -4.9 $\pm$ 2.4 \\
60466.0196  & 0.0152 $\pm$ 0.0014 & -0.0030 $\pm$ 0.0014 & 1.57 $\pm$ 0.14 & 0.3 $\pm$ 2.5  \\
60466.0225  & 0.0108 $\pm$ 0.0014 &  0.0023 $\pm$ 0.0014 & 1.11 $\pm$ 0.14 & -2.5 $\pm$ 3.5 \\
60466.0253  & 0.0106 $\pm$ 0.0014 & -0.0037 $\pm$ 0.0014 & 1.19 $\pm$ 0.13 & -7.3 $\pm$ 3.3 \\
\hline
\end{tabular}
}
\end{table}

\begin{figure}
\begin{center}
\includegraphics[width=0.48\textwidth]{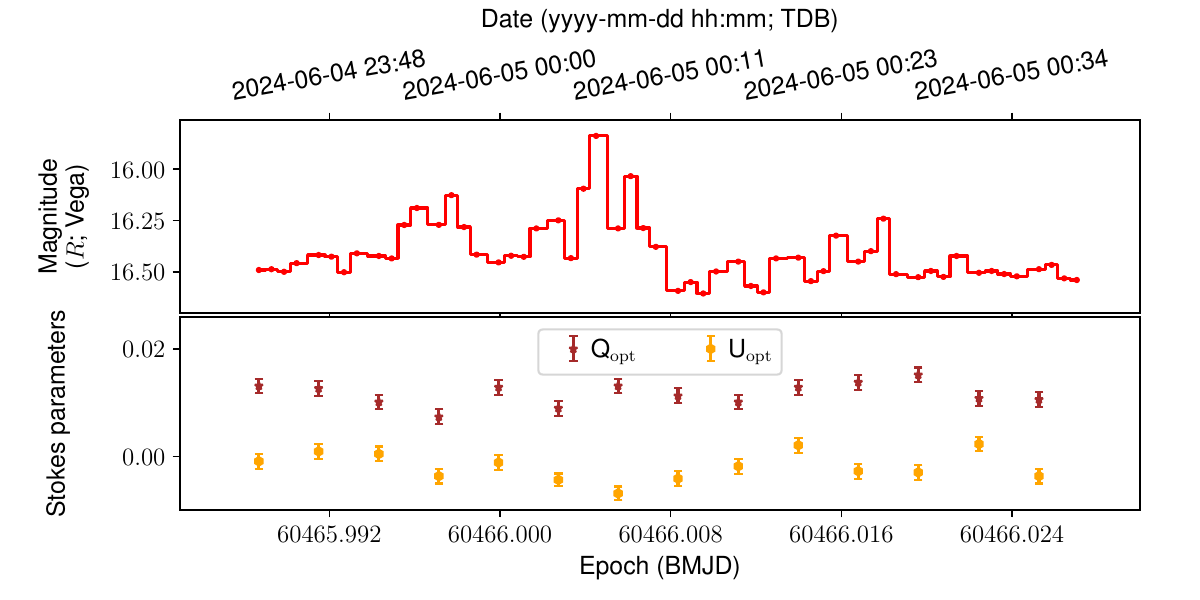}
\includegraphics[width=0.48\textwidth]{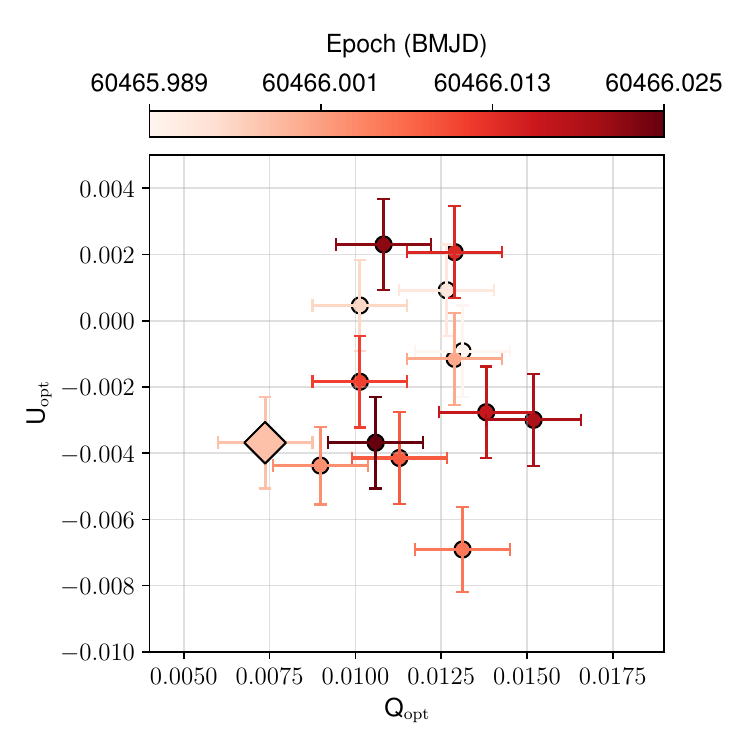}
\vspace{-0.8cm}
\caption{Top: time evolution of optical intensity (red) and of the Q$_{\rm opt}$ (brown) and U$_{\rm opt}$ (orange) Stokes parameters. Bottom: Scatter plot of Stokes parameters for the optical emission from J1023. Data points are color-coded based on the time of their measurements, with a diamond marker highlighting the epoch corresponding to the low mode detected during simultaneous X-ray observations. BMJD stands for Barycentric Modified Julian Date. TDB indicates the Barycentric Dynamical Time.}
\label{fig:Stokes_opt_ev}
\label{fig:Stokes_opt_plane}
\end{center}
\end{figure}

\section{Polarization Properties in the Low Mode; Constraints on Polarized Radio Emission in the two Modes}
\label{sec:LM_app}
In the scenario outlined in Section\,\ref{sec:mini-pwn}, the upper limit on the radio linear polarization fraction in the high mode is consistent with partially self-absorbed synchrotron radiation from the optically thick portion of the compact jet.
The exact cause of the switch to the low mode remains uncertain, though the detection of short mm-band flares during a few of these switches in past observations points to sudden ejections of magnetized plasma, likely originating from the innermost regions of the accretion flow. The ejection of this matter could thus force the boundary region to move outward, reducing the X-ray flux and the pulsed emission. As the plasma propagates downstream in the compact jet, it becomes optically thin in the radio band \citep{Baglio2023}. 
This framework also accounts for the optical polarization properties measured during the low mode. If the boundary region moves outward, it may still contribute a measurable amount of polarized light. Moreover, the optical polarization angle observed in the low mode is consistent with that of the high mode within uncertainties, suggesting a common underlying emission mechanism for the optical linear polarization across both modes and reinforcing the idea that residual emission from the collision remains active in the low mode. In the X-ray band, where the jet could dominate up to 100\% of the emission \citep{Baglio2023}, the maximum expected polarization would be $\simeq$25\%. The upper limit for the X-ray polarization degree during the low mode, $P_{\rm X,L} < 26$\% at a 90\% c.l., makes it impossible to draw firm conclusions from this measurement.

Alternatively, the switch to the low mode may result from matter occasionally penetrating the light cylinder, temporarily forcing the system into a regime where the rapidly rotating magnetic field anchored to the NS propels the inflowing matter away. As matter expands outward, it becomes increasingly rarefied and more transparent to synchrotron emission at radio wavelengths \citep{Veledina19}.

At radio wavelengths, synchrotron radiation is expected to produce linear polarization fractions of up to 10\% for optically thick emission and up to 70\% for optically thin emission, assuming uniform magnetic fields and a power-law electron distribution with an index of $\simeq$2 \citep{longair11}. In accreting stellar-mass black holes, polarization fractions are typically below 10\%, but can reach up to 50\% during instances of optically thin emission (\citealt{Hughes2023} and references therein). 




\end{document}